# Understanding the liquid jet break-up in various regimes at elevated pressure using a compressible VOF-LPT coupled framework

Bharat Bhatia[1], Tom Johny[1], and Ashoke De[1,2*]

[1] Department of Aerospace Engineering, Indian Institute of Technology Kanpur, 208016, Kanpur, India.

[2] Department of Sustainable Energy Engineering, Indian Institute of Technology Kanpur, 208016, Kanpur, India.

## Abstract

The present work develops a compressible Volume of Fluid (VOF) – Lagrangian Particle Tracking (LPT) coupled solver in OpenFOAM and utilizes it to simulate a liquid jet in crossflow (LJICF) numerically. This methodology helps accurately predict a complex primary breakup in the Eulerian framework and the secondary atomization of spherical droplets using a computationally efficient LPT method. The coupled solver with Adaptive Mesh Refinement (AMR) is rigorously validated for a liquid jet in crossflow at varying operating conditions – pressure, crossflow velocity, and inlet liquid jet velocity. We have further carried out a thorough investigation to study the effect of momentum flux ratio and weber number on the various flow features and liquid jet break-up phenomenon in a crossflow while identifying the stream-wise location of the liquid jet breakup region. At low momentum flux ratios in the bag breakup regime, the predictions reveal that the liquid jet breakup occurs due to the growth of similar instability as usually observed in the high-speed liquid sheet atomization. The short wavelength assumption of the inviscid dispersion relation resembles the Kelvin-Helmholtz type instability observed in this case, as opposed to Rayleigh-Taylor instability at high momentum flux ratio in the surface breakup regime. It is also proposed that the shear breakup along the transverse edges of the liquid column occurs due to the shear layer instability of the air passing around the liquid column. The simulation wavelength closely matches the Williamson correlation for shear layer instability around cylinders – a shape similar to the cross-section of the bottom of the liquid column. The results show a distinct streamer or bifurcation phenomenon at low momentum flux ratios and moderate weber numbers. Further investigation suggests that the internal liquid boundary layer and the three-dimensional flow field behind the liquid jet are responsible for streamer formation.

# 1. Introduction

Spray atomization of a liquid jet in crossflow is a complex multiphase phenomenon extensively used in aviation engines, lean premixed prevaporized (LPP) ducts, film cooling of turbine blades, rockets, scramjets, augmentors, and other applications. With the new environmental restrictions and regulations to reduce NOx emissions becoming more stringent, optimizing the combustion process in these systems has become critical. Transverse fuel injection into a crossflow enhances combustion efficiency, cuts fuel consumption, and lowers emissions in such systems. As a result, to develop and analyze such systems, a detailed understanding of the physics of liquid jet break-up and atomization processes is required. Experimental investigations of the spray atomization process require precise and expensive experimental techniques. The numerical studies can provide valuable insight into atomization, especially when flow features close to nozzle regions are difficult to capture experimentally. Multiphase flow modeling is inherently complex due to the wide range of scales formed and the liquid-gas phase interaction resulting in complex structures and flow patterns. The length scale here ranges from a few micrometers to several centimeters.

A liquid jet in a crossflow (LJICF) consists of two significant stages of break-up and atomization: primary and secondary breakup. (1) Primary break-up results from the instabilities at the liquid-gas interface, which grows in size due to the inertial forces and turbulence in the liquid jet, causing the liquid core to break up into large liquid structures. (2) Secondary break-up further breaks these liquid structures and ligaments into smaller droplets. The extent to which these droplets are disintegrated is influenced by the external distorting forces and surface tension forces of the liquid, finally resulting in a large number of stable spherical droplets. The break-up of a liquid jet in a cross-flow is schematically represented in Figure 1. The crossflow atomization process involves complex flow features, mainly the turbulent break-up, droplet deformation from the liquid-gas interaction, vortex formations, etc. It also involves mass, momentum, and energy exchange between the liquid and gaseous phases.

Most of the early studies of spray atomization for LJICF were primarily experimental, where researchers mainly concentrated on the liquid jet penetration, break-up and its different modes, and their relationship with various parameters, especially the momentum flux ratio (q) and crossflow Weber number (We) (Wu et al., 1997; Becker and Hassa, 2002; Sallam et al., 2004; 2006). Wu et al. (1997) studied the liquid jet penetration for various test liquids (water, alcohol, and their mixtures) under different operating conditions and plotted a transition regime between column and surface break-up modes. They also found that the column fracture location in the streamwise direction is constant and is equal to 8D downstream of the nozzle exit. Stenzler (2006) proposed that the aerodynamic weber number and liquid viscosity can also affect the jet penetration and the momentum flux ratio. Ingebo (1957; 1967) also proposed the same by considering the effect of liquid viscosity in the spray penetration. He also noticed that larger droplets penetrate deeply into the cross-flow and affect the penetration. As a result, he proposed a spray trajectory correlation that includes the effect of larger droplets by taking account of parameters ($Re_{jet}$, We) other than the momentum flux ratio (q). Several others looked into the various break-up modes, including Becker and Hassa (2002), who proposed a qualitative and visual map for break-up and atomization. M. Eslamian (2014) proposed a similar map based on the momentum flux ratio and crossflow weber numbers for the primary break-up and a transition line between column and shear break-up modes. Madabhushi et al. (2004) redefined the break-up regime based on the turbulent transition jet Reynolds number and cross-flow Weber number. Since the momentum flux ratio does not provide the turbulence details of a liquid jet, they defined a borderline based on the Reynolds number of the jet in addition to the break-up map to include the effect of liquid jet turbulence. Sallam et al. (2006) classified the break-up mechanisms for non-turbulent liquid jets into four based on the aerodynamic weber number: column, bag, multimode, and shear break-up. There are few studies available in the literature regarding the droplet size characteristics.

Several computational techniques have been developed to simulate the spray atomization process numerically. Computing interface motion in complex multiphase flows such as spray atomization requires accurate interface tracking and reconstruction techniques. The interface capturing approaches such as the Volume of Fluid (VOF) (Hirt and Nichols, 1981), Front tracking methods (Tryggvason et al., 2001) or CLSVOF (Sussman et al., 2000; 2013; Menard et al., 2007) are typically used modeling approaches for the primary breakup. VOF methods are based on the volume fraction (α) of phases in a computational cell. An additional transport equation is solved for volume fraction (α), which is then used to track the location of the liquid-gas interface. Many VOF interface capturing techniques are utilized to capture the interface, classified as algebraic or geometric. Older methods for interface capture include algebraic approximations such as Compressive schemes and THINC (Tangent of hyperbola for interface capturing) schemes. Geometric approaches are more recent, complicated, and precise (Mirjalili, S. 2019). An interface within a computational cell is geometrically reconstructed using a plane in three-dimensional (3D) simulations using methods such as Simple Line Interface Calculation (SLIC) (Noh & Woodward, 1976) or the more recent Piecewise Linear Interface Calculation (PLIC). Nonetheless, using these techniques for commercial use does have a disadvantage: high computational cost and time. The grid resolution must be good enough in a VOF approach to sufficiently resolve the liquid structures, droplets, and ligaments. Therefore, the computational cost of simulating a complete spray break-up using the only VOF method is very high (Heinrich and Schwarze, 2020). The Lagrangian Particle Tracking (LPT) methods are more suitable for secondary atomization involving a cloud of dispersed droplets, owing to lower computational cost and better droplet predictions. Thus, the VOF approach can efficiently resolve the primary break-up, while the dispersed cloud uses the LPT method. Therefore, an Eulerian-Lagrangian coupled approach could significantly reduce the computational cost of simulating the whole spray atomization process while still capturing the underlying physics to a high degree.

A coupling algorithm is an intermediate between the Eulerian and Lagrangian frameworks, which tracks all the Eulerian-phase droplets throughout the computational domain and replaces them with a Lagrangian substitute using transformation criteria. Several coupling algorithms (Hermann et al., 2008; 2010; H. Grosshans, 2014; Heinrich and Schwarze, 2020; H. Yu et al., 2017) have been proposed for the Eulerian-Lagrangian droplet transformation. H. Grosshans (2014) performed a statistical coupling approach by defining a coupling layer between VOF and LPT frameworks. The drawback associated with this method is that the 2-D coupling layer is fixed in space. Also, the shearing action generates smaller droplets from the sides of the liquid column. Since these droplets lie before the coupling layer, they are resolved using VOF only, even though these are more appropriate to be tracked using LPT. So the transformation into Lagrangian droplets may not be realistic as it seems. H. Yu et al. (2017) proposed a region coupling method (RCM) with a droplet identification and extraction technique. This method is more realistic since the transformation happens smoothly in the coupling region defined in three-dimensional (3-D) space. One drawback of this method is the placement of the coupling region, which has to be determined previously from another Eulerian simulation. Hermann et al. (2008; 2010) used their band generation algorithm to couple the Lagrangian framework with a refined level-set grid method. Heinrich and Schwarze (2020) employed an image processing algorithm called Connected Component Labelling (CCL) to couple the VOF with LPT for incompressible flows. Such a coupled methodology eliminates solutions through stochastic methods such as the ELSA model (Hoyas et al., 2013) or other models that use a Lagrangian droplet ejection out of a primary jet core calculated using the VOF method (Saeedipour et al., 2016), ignoring the critical phenomenon of the primary breakup.

One assumption involved in the LPT method is that the droplet volume is minimal compared to the local cell volume. For the LPT approach to be numerically stable, it is generally advised that the grid size be larger by ten times the size of the droplets (Vallier et al., 2011). While Arlov et al. (2007) proved that the LPT theory is valid even if the cell size is more than the droplet size by five times. Multiple grid cells comprising the Eulerian droplet are required for an Eulerian framework to resolve the small-scale features adequately. To overcome the imbalance of larger grid size requirement for the LPT method and smaller grid size for VOF, we can either use: the (i) Adaptive Mesh Refinement (AMR) technique based on the liquid volume fraction ($\alpha$) or (ii) a static grid with a highly refined region for Eulerian framework and a separate coarser grid for the lagrangian framework (Herrmann 2010). The former option is preferable as the latter may cause larger lagrangian droplets on a smaller local mesh volume.

In this work, a compressible VOF-LPT coupled solver is developed in OpenFOAM along with the evaporation models for Eulerian and Lagrangian fields, which is also capable of simulating atomization involving high temperature evaporating sprays. It uses a VOF-LPT coupling algorithm based on the previous works of Heinrich and Schwarze (2020). The liquid-gas interface is reconstructed geometrically using the isoAdvection concept developed by Roenby et al. (2016). Such a coupled model and an additional method for interface capturing would help capture the flow physics to a high degree in the near-nozzle region and the far downstream region while keeping the computational cost to a minimum. The developed model is validated under elevated pressure and room temperature conditions for a liquid jet-in-crossflow case based on the experimental works of Amighi and Ashgriz (2019) in terms of droplet size characteristics and the Sauter mean diameter, the standard deviation of droplets, and the jet trajectory. And further, we have carried out a detailed investigation using the same framework for a wide range of parameters (q, We) and its effect on spray droplet sizes ($D_{32}$, STD), liquid jet penetration, and other flow features (vortex formation, break-up behavior, and deformation) for a liquid JICF under different operating conditions.

## 2. Methodology

This section describes the methodologies employed in our current investigation, mainly the VOF Eulerian formulation, the LPT formulation, and the coupling algorithm.

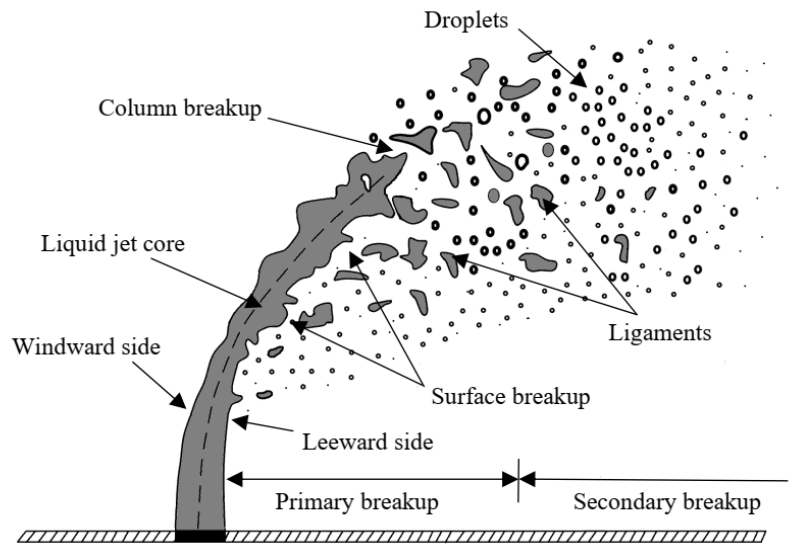

Fig. 1: Schematic of the break-up of a Liquid Jet in Crossflow (LJICF)

## 2.1 Eulerian framework

The VOF method tracks the interface between the two phases in the Eulerian framework. The mass, momentum, species mass fraction, and energy equations are solved for a two-phase compressible and immiscible system. VOF method uses the volume fraction $\alpha$, defined as a step function, to distinguish between the two phases. A volume fraction value of $\alpha_1 = 1$ indicates cell volume fully occupied by liquid, and $\alpha_1 = 0$ indicates cell volume fully occupied by gas. A volume fraction value $0 < \alpha_1 < 1$ shows the liquid-gas interface within the cell control volume.

$$\alpha_1 = \begin{cases} 0 & phase2 \\ 0 < \alpha_1 < 1 & interface \\ 1 & phase1 \end{cases} \tag{1}$$

The step function α makes it possible to solve only one set of governing equations for both phases, eliminating the need for separate equations for each phase. The VOF method without any interface reconstruction method results in smearing the liquid surface. To overcome the smearing of the profile at the interface, many interface-capturing methods have been developed for VOF involving the reconstruction of the interface. Reconstruction methods may be either geometric or algebraic – like the piecewise-linear interface calculation (PLIC) for geometric reconstruction and Compressive Interface Capturing Scheme for Arbitrary Meshes (CICSAM) for algebraic reconstruction in Ansys Fluent (ANSYS, Inc., 2016) or Gerris solver (Popinet, 2003), Multidimensional Universal Limiter with Explicit Solution (MULES) scheme for algebraic reconstruction by interFoam solver in OpenFOAM (Weller et al., 1998), etc. The volume fraction $\alpha_k$ and the interface within a cell is reconstructed geometrically using the iso-advection concept devised by Roenby et al. (2016). For geometric interface reconstruction, it uses a piecewise linear interpolation calculation (PLIC) (Mencinger et al., 2011). Although the method uses an integral form of the continuity equation to calculate the surface evolution, it is represented here in the differential form to maintain consistency with the rest of the equations:

$$\frac{\partial(\rho)}{\partial t} + \nabla.(\rho\mathbf{u}) = S_\rho\big|_L \tag{2}$$

Here $\rho$ is the mixture density, $\mathbf{u}$ is the velocity, $S_\rho\big|_L$ is the source terms from the Lagrangian droplets calculated as:

$$S_\rho|_L = 4\pi \frac{k}{c_p} r_0 \frac{1}{1+G_f/G} ln\left[1 + \frac{h_g - h_{drop,s}}{L(T_b)}\left(1 + \frac{G_f}{G}\right)\right] \tag{3}$$

The source term denotes the heat transfer vaporization rate (Zuo et al. (2000)); $k$ is the thermal conductivity of gas, $c_p$ is the specific heat of the gas; $h_g$ and $h_{drop,s}$ denotes the enthalpy of the gas and at the droplet surface, respectively; $L(T_b)$ is the latent heat of vaporization at the boiling temperature, $G_f$ is the flash-boiled vapor mass flow rate, which reduces to zero at temperatures less than or equal to the boiling point. In the integral form of the equation, the instantaneous rate of change of the total mass within a volume is equal to the instantaneous flux of mass through its boundary in addition to the mass evaporated from the lagrangian particles. In the differential form, the mass conservation equation for phase i is:

$$\frac{\partial(\alpha_k \rho_k)}{\partial t} + \nabla.(\alpha_k \rho_k \mathbf{u}) = S_\rho\big|_L + S_\rho\big|_E \tag{4}$$

While the source term from liquid to the gas phase, it is:

$$S_\rho|_E = 2 * (1 - \alpha_2) * \mathrm{M}_{tc} * \mathrm{D}_{ab} * dY. \tag{5}$$

Here, $\mathrm{M}_{tc}$ and $\mathrm{D}_{ab}$ are the mass transfer and diffusion coefficients between two phases, and $dY$ is the species gradient near the interface for liquid species. While the volume fractions are constrained, this equation evolved in two steps. Firstly, the reconstruction step is where the distribution of fluids inside the computational cells is estimated using an efficient iso-surface calculation methodology. Secondly, a face-interface intersection line sweeping the face for a sub-time interval approximates the time evolution of the submerged part of a general polygon face (belonging to a computational cell). The sub-time interval is defined by the time a line passes the face vertices. It makes the analytical calculation possible for the passage of a fluid across the cell face during this sub-time interval. This reconstruction technique applies to structure as well as unstructured grids.

For cells having liquid-gas calculations, the fluid properties are calculated from the weighted average of the phase fraction α for each computational cell.

$$\rho = \alpha_1 \rho_1 + \alpha_2 \rho_2 \tag{6a}$$

$$\mu = \alpha_1 \mu_1 + \alpha_2 \mu_2 \tag{6b}$$

where $\mu_1$, $\mu_2$ and $\rho_1$, $\rho_2$ are the dynamic viscosity and density of phases 1 and 2, respectively. Phases 1 and 2 represent the liquid and gas phases for a two-phase compressible system. A single momentum transport equation is solved for the velocity field in both phases:

$$\frac{\partial(\rho u)}{\partial t} + \nabla.(\rho \mathbf{u}\mathbf{u}) - \nabla \tau_{eff} = -\nabla p_d - (\nabla \rho)\mathbf{g}.\mathbf{x} + F_{ST} + S_{\rho u}\big|_L + S_{\rho u}\big|_E \qquad (7)$$

where the piezometric pressure $p_d = p - \rho \mathbf{g}.\mathbf{x}$ and $\tau_{eff} = \mu_{eff}\left[\left(\nabla \mathbf{u} + \nabla^T \mathbf{u}\right) - \frac{2}{3}tr(\nabla \mathbf{u})\right]$; here, $\mu_{eff}$ the effective viscosity is volume averaged as: $\mu_{eff} = \frac{4(\alpha_1 \mu_1 + \alpha_2 \mu_2)}{3}$, and $\rho$ in all the combined equations is the mixture density for each computational cell. $S_{\rho u}\big|_L$, $S_{\rho u}\big|_E$ are the Lagrangian and Eulerian source terms for momentum equations, which are generated because of the atomization and the evaporation of droplets in the domain. $F_{ST}$ is the surface tension force calculated using the Continuum Surface Force (CSF-model) formulation proposed by Brackbill et al. (1992). The surface tension force is defined at the interface where a pressure jump occurs and is specified as a source term to the momentum equation. It is calculated as follows:

$$\vec{F}_{ST}(\vec{x}) = \sigma \int_S \kappa(\vec{y})n(\vec{y})\delta(\vec{x} - \vec{y})dS \qquad (8)$$

where $\sigma$ is the surface tension of the liquid phase, $\kappa$ is the local interface curvature, $n$ is the unit interface normal, and $\delta$ the Dirac-delta function. $\vec{x}$ and $\vec{y}$ are the position vectors where the forces are calculated. These are calculated using: $\kappa = -\nabla.n$ , $n = \frac{\vec{n}}{|\vec{n}|}$ where $\vec{n}$ = normal surface vector, defined as $\vec{n} = \nabla \alpha$. The surface tension force source term can be expressed as:

$$F_{ST} = \sigma \frac{\rho \kappa \vec{\nabla} \alpha}{\frac{1}{2}(\rho_l + \rho_g)} \qquad (9)$$

The equation of state is used to solve densities from pressure and temperature conditions. Considering an isentropic gas phase, the equation of state is defined as (Miller et al., 2013):

$$\frac{p}{\rho^\gamma} = a_c = \text{constant} \qquad (10)$$

where $a_c$ is the isentropic constant and $\gamma$ is the ratio of specific heat. The total derivative of $\rho$ with respect to pressure gives:

$$\left(\frac{\partial \rho}{\partial p}\right)_s = \frac{1}{a_c \gamma}\left(\frac{p}{a_c}\right)^{\frac{(1-\gamma)}{\gamma}} \qquad (11)$$

The speed of the sound wave in a liquid medium is computed as follows ([Miller et al., 2013](#)):

$$\left(\frac{d\rho}{dp}\right)_s = \frac{1}{c^2} \tag{12}$$

where c is the speed of the sound. Integration of the above equation at a constant speed of sound yields:

$$\rho - \rho_o = \psi(p - p_o) \quad , \text{ where } \quad \psi = \frac{1}{c^2} \tag{13}$$

and $\rho_o$, $p_o$ are the reference density and pressure, respectively.

The species mass fraction equation is solved by assuming the liquid phase as a single component system and the gaseous phase as a multi-component system.

$$\frac{\partial(\alpha_2\rho_2 Y_k)}{\partial t} + \nabla.(\alpha_2\rho_2\mathbf{u}Y_k) - \nabla.(\alpha_2 D_k\nabla Y_k) = S_{\rho Yk}\big|_L + S_{\rho Yk}\big|_E \tag{14}$$

Here, $Y_k$ is the species mass fraction and $D_k$ is the diffusion coefficient for the kth species for the gaseous phase. The energy equation is also solved since our system is compressible.

$$\frac{\partial(\rho T)}{\partial t} + \nabla.(\rho\mathbf{u}T) - \nabla.(\alpha_{T,eff}\nabla T) + \nabla.(\mathbf{u}p) + \left(\frac{\partial(\rho K)}{\partial t} + \nabla.(\rho\mathbf{u}K)\right)\left(\frac{\alpha_1}{c_{v,1}} + \frac{\alpha_2}{c_{v,2}}\right) = S_{\rho T}\big|_L + S_{\rho T}\big|_E \tag{15}$$

Here, T is the temperature in the Eulerian frame at any location, $\alpha_{T,eff}$ is the effective thermal diffusivity, $K$ is the kinetic energy ($K = 0.5 * |\mathbf{u}|^2$), $c_{v1}$ and $c_{v2}$ are specific heat capacities at constant volume for phases 1 and 2, respectively. $S_{\rho T}|_L$, $S_{\rho T}|_E$ are the lagrangian and eulerian source terms for the energy equation, which accounts for the atomization and the evaporation of droplets. Although evaporation does not occur in our low-temperature test conditions, the developed framework accounts for the effect of droplet evaporation in both the Eulerian and Lagrangian frameworks.

Large-eddy simulations (LES) are used to model turbulence in the flow field. The large eddies of the turbulent flow are computed directly in LES. Sub-grid scale (SGS) modeling is performed since the dissipative scales of turbulence are not entirely resolved in LES. The SGS models the effect of small-scale vortices and eddies on the resolved larger eddies. Thus, the SGS terms cannot be calculated and require closure models. The SGS Reynolds stress ($\tau_{ij}$) and the SGS heat flux ($Q_j$) are the parameters that require closure models at the sub-grid scales, as given by

$$\tau_{ij,SGS} = \overline{\rho u_i u_j} - \overline{\rho} \, \overline{u_i} u_j \tag{16}$$

$$Q_{j,SGS} = \overline{\rho u_j h} - \overline{\rho} \, \overline{u_j} h \tag{17}$$

and the molecular strain rate tensor is given by:

$$\overline{S}_{ij} = -\frac{2}{3}\mu \frac{\partial u_k}{\partial x_k}\delta_{ij} + \mu\left(\frac{\partial u_i}{\partial x_j} + \frac{\partial u_j}{\partial x_i}\right) \tag{18}$$

The dynamicKEqn model, which is a one-equation eddy viscosity SGS model, is used for the present study. This model was primarily developed by Kim and Menon (1995) based on the previous works of Germano et al. (1991). In recent years, many improvements to the models have been proposed (Chai and Mahesh, 2012, Huang and Li, 2010). The dynamicKeqn is represented as follows:

$$\frac{\partial k_{sgs}}{\partial t} + \frac{\partial U_j k_{sgs}}{\partial x_j} = \frac{\partial}{\partial x_j}\left(\nu_t \frac{\partial k_{sgs}}{\partial t}\right) - C_e \frac{k_{sgs}^{3/2}}{\Delta} - 2\nu_{sgs} S_{ij} S_{ij} \tag{19}$$

$$\nu_{sgs} = C_k \sqrt{k_{sgs}} \Delta \tag{20}$$

where $k_{sgs}$ is the sub-grid scale kinetic energy, $\nu_t$ is the effective kinematic viscosity (both molecular and sub-grid viscosity). The $C_k, C_e$ constants are computed based on the dynamic formulation from Germano et al. (1991).

The partial differential equations (PDEs) are discretized using a Finite Volume Method (FVM) code implemented in the OpenFOAM v1912. The convective flux discretization deploys a second-order TVD (Total Variation Diminishing) scheme, while the viscous flux discretization involves a second-order central scheme, and the temporal term deploys the first-order implicit Euler scheme with sufficiently small time steps to maintain stability and reduce numerical diffusion.

## 2.2 Lagrangian framework

In the LPT method, the liquid droplets are treated as point particles with mass and momentum but no volume. The LPT approach uses parcels to represent a group of droplets with similar characteristics such as droplet size, velocity, temperature, and thermophysical properties. The particle position and velocities are updated at each time step using the Basset Boussinesq-Oseen (BBO) equation (Parmar et al., 2011):

$$\frac{d\mathbf{x}_p}{dt} = \mathbf{u}_p \tag{21}$$

$$m_p \frac{d\mathbf{u}_p}{dt} = \mathbf{F}_D + \mathbf{F}_G \tag{22}$$

where $\mathbf{x}_p$, $m_p$, $\mathbf{u}_p$ is the position, mass, and velocity of each particle, respectively. $\mathbf{F}_D$ and $\mathbf{F}_G$ are the drag and gravitational forces (body forces) acting on the particles. They are calculated as follows:

$$\mathbf{F}_D = C_D \frac{\pi D_p^{\,2}}{8} \rho_g (\mathbf{u}_g - \mathbf{u}_p) \,|\, \mathbf{u}_g - \mathbf{u}_p \,| \tag{23}$$

$$\mathbf{F}_G = m_p g \left( 1 - \frac{\rho_g}{\rho_p} \right) \tag{24}$$

$\mathbf{F}_G$ account for both gravity and buoyancy effects, $D_p$ is the diameter of the lagrangian parcel, $\mathbf{u}_g$ and $\mathbf{u}_p$ are the velocities of gas-phase and velocity of Lagrangian parcels. $C_D$ is the drag coefficient of droplets, which is calculated from the Schiller-Naumann equation (Schiller, 1935):

$$C_D = \begin{cases} 24(1 \ + \ 0.15 Re_p^{\,0.687})/Re_p & Re_p \leq 1000 \\ 0.44 & Re_p > 1000 \end{cases} \tag{25}$$

$$\mathrm{Re}_p = \frac{\rho_2 \left| \mathbf{u}_g - \mathbf{u}_p \right| D_p}{\mu_2} \tag{26}$$

where $Re_p$ is the Reynolds number of the lagrangian droplets and $\mu_2$ is the gas phase dynamic viscosity.

The LPT is intended to characterize every droplet feature like droplet break-up, heat transfer, and evaporation. The Ranz-Marshal model (Ranz and Marshall, 1952) is used for heat transfer calculations. Since the Lagrangian droplets are assumed to be spherical, the evaporation is calculated based on the Frossling correlation (Frössling, 1938):

$$Sh = 2 + 0.552 Re_p^{\,1/2} Sc^{1/3} \tag{27}$$

$Sh$ is defined as the ratio $h_c d_p / D$, where $h_c$ denotes the convective mass transfer coefficient, and $d_p$ is the droplet diameter, $Re_p$ is the droplet Reynolds number, and $Sc$ is the Schmidt number. The typical value of the turbulent Schmidt number (Prandtl number) ranges from 0.7 to 1. A value of 0.85 was used in this work.

Among the atomization models of TAB (O'Rourke & Amsden, 1987), ETAB (Tanner, 1997), and R-D (Reitz and Diwakar, 1997), the Reitz-Diwakar secondary break-up model is chosen to model the break-up of the parcels due to the aerodynamic forces. This is suitable for our high-pressure test conditions and resulted in better droplet predictions. This takes into account the bag and the stripping break-up of droplets. A critical Weber number is used to dictate the breakup of droplets. Droplet collision is an important parameter that mainly affects droplet numbers, droplet sizes, droplet evaporation, spray propagation, and distribution. To account for the collision and coalescence of the droplets, we employ the trajectory model of Schimdt & Rutland (2000). A trajectory-based collision model is more realistic as it models droplet collisions using droplet position and velocity vectors.

## 2.3 Coupling algorithm

As mentioned in the previous sub-sections, we have developed the compressible VOF-LPT coupled solver in the OpenFOAM framework along with the evaporation models for Eulerian and Lagrangian fields which is also capable of simulating atomization involving high-temperature evaporating sprays. The algorithm provided by Heinrich & Schwarze (2020) is used to couple the compressible VOF framework with LPT and facilitate the droplets' transformation from one framework to another. This coupling algorithm first identifies the individual droplets present throughout the three-dimensional (3-D) domain in the Eulerian framework. The relevant properties are then calculated, and if the droplets satisfy the transformation criteria, they are injected as Lagrangian parcels after deleting the corresponding liquid droplet from the Eulerian framework. An image processing – Connected Component Labelling (CCL) algorithm is used in the methodology to identify various liquid droplets in the Eulerian domain (Heinrich & Schwarze, 2020). The liquid volume fraction ($\alpha_1$) is used to scan the complete domain for the connectivity of liquid-filled cells. Two separate, detached lumps have cells $\alpha_1 < \alpha_{\min}$ in between them, where $\alpha_{\min}$ is the minimum value of liquid volume fraction to distinguish two separate droplets. The details of the algorithm have been explained in Heinrich & Schwarze (2020).

The properties of each Eulerian droplet are then calculated as follows:

$$V_p = \sum_i \alpha_i V_i \tag{28a}$$

$$\boldsymbol{x}_p = \frac{1}{V_p} \sum_i \alpha_i V_i \boldsymbol{x}_i \tag{28b}$$

$$\boldsymbol{u}_p = \frac{1}{V_p}\sum_i \alpha_i V_i \boldsymbol{u}_i \qquad (28c)$$

$$T = \frac{1}{V_p}\sum_i \alpha_i V_i T \qquad (28d)$$

$$d_p = \sqrt[3]{\frac{6V_p}{\pi}} \qquad (28e)$$

All the cells belonging to a particular Eulerian droplet are represented by the index $i$. The equivalent diameter ($d_p$) is the diameter of a sphere with the volume $V_p$ equal to that of the Eulerian droplet, whereas $\boldsymbol{x}_p$, $\boldsymbol{u}_p$ and $T$ are the position, velocity, and temperature of the Eulerian droplet, respectively. Lastly, the transformation criteria for each droplet are reviewed. If the droplets meet the transformation criteria, they are injected as Lagrangian parcels, preserving their velocity, momentum, energy, and position. And the corresponding Eulerian liquid droplet is deleted from the Eulerian framework. The transformation criteria are based on the assumption of the size of droplets and their shape. Firstly, the size of the droplet diameter ($D_p$) should be smaller than a *predetermined minimum diameter* because a larger droplet is more prone to deformation, which is better resolved in the Eulerian framework. The minimum diameter criteria refer to the droplet size below which they are eligible for conversion into Lagrangian droplets. This is also determined from the experiments where the value used is 280µm (Amighi, 2015). On the other hand, smaller droplets are converted as they cannot be appropriately resolved in the Eulerian framework and are better tracked in the Lagrangian framework. Secondly, if the *sphericity* of the droplet is less than a threshold minimum.

In this study, we have used a sphericity of 1.47, calculated based on the circularity of droplets reported in the experiments (Amighi and Ashgriz, 2019). Droplets with sphericity less than 1.47 will be converted to lagrangian droplets, while those with sphericity more than 1.47 will remain as Eulerian droplets. Heinrich & Schwarze, (2020) used a sphericity of 2 in their work. It is desirable for the droplets with sphericity greater than the threshold to stay in the Eulerian framework as the irregular droplets are vulnerable to deformation, which is well captured by the Eulerian method only. Here we have performed a one-way coupling that facilitates the transformation of Eulerian droplets into Lagrangian only, while a two-way coupling also does vice-versa. The one-way coupling assumes only minor differences between the couplings, as observed in Ling et al. (2015) and Fontes et al. (2019).

## 2.4 Computational domain and flow conditions

The above-proposed model is validated using experimental data from Amighi and Ashgriz (2019) at elevated pressure conditions for an experimental channel. Figure 2 illustrates the computational domain with the same cross-section of $25\times35$ mm$^2$ (y×z) as the experimental channel. The computational domain in this study is 58.6mm in length (x-direction), which is only a part of the experimental channel size. Following the experimental measurements, D$_{32}$, STD calculations are done within a 52.45D distance downstream (blue sub-domain) from the point of liquid jet injection (28.6, 12.5, 0) mm. The liquid jet is injected from the nozzle placed at the bottom wall of the domain in the z-direction (Jet direction). The exit diameter of the nozzle used is 572 µm.

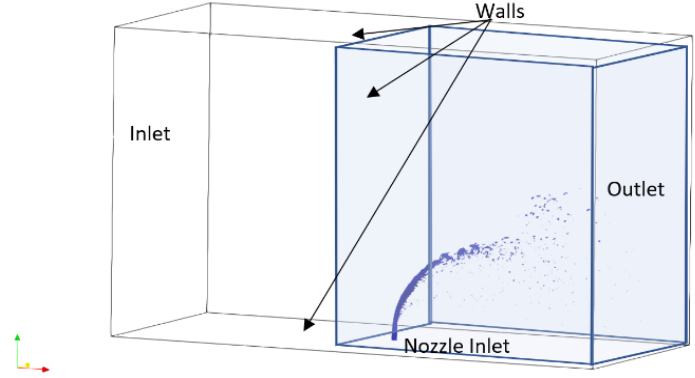

Fig. 2: Computational domain and region of calculation

The initial domain meshes with a $140 \times 60 \times 80$ cells, and the cells are additionally refined around the injector (see Figure 3). We have performed four levels of refinement, with the finest grid size being 26 µm in the nozzle region to the coarsest level of 418 µm of the parent mesh. This results in a total size of 0.75 million cells in the parent mesh.

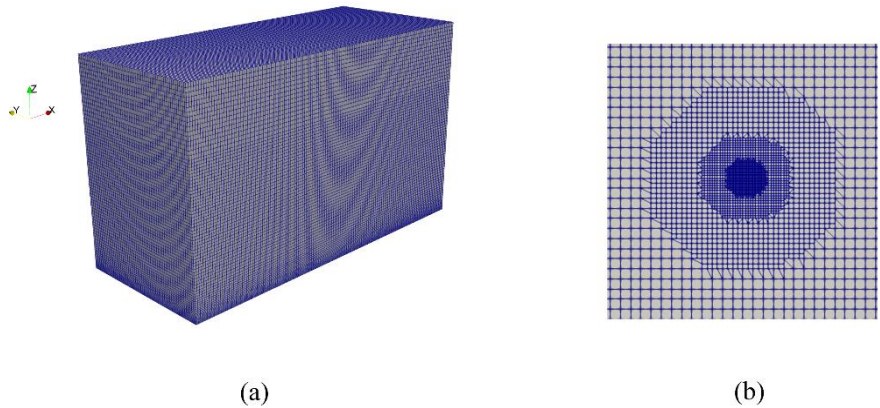

(a)                                      (b)

Fig. 3: Mesh generated with (a) blockMesh and (b) SnappyHexMesh (near nozzle) with four levels of refinement near the nozzle

On our parent mesh, we have used adaptive mesh refinement (AMR) with three refinement levels to accurately refine the liquid-gas interface region and capture the complicated dynamics of the interface. A lower and upper refine level values ($0.1 < \alpha_1 < 1$) of the volume fraction alpha are used as a criterion to refine the cells having a liquid-gas interface. This continues refining the cell until the maximum refinement is achieved or if the alpha values fall outside the acceptable range. The lower refine level value of 0.1 is chosen after checking for different refinement levels as the lower values (<0.1) yield similar results. The use of AMR reduces the computational requirements considerably; thus, a coarser mesh can be employed. With the help of AMR, the mesh resolution of 52 μm is achieved just at the interface (refer Figure 4).

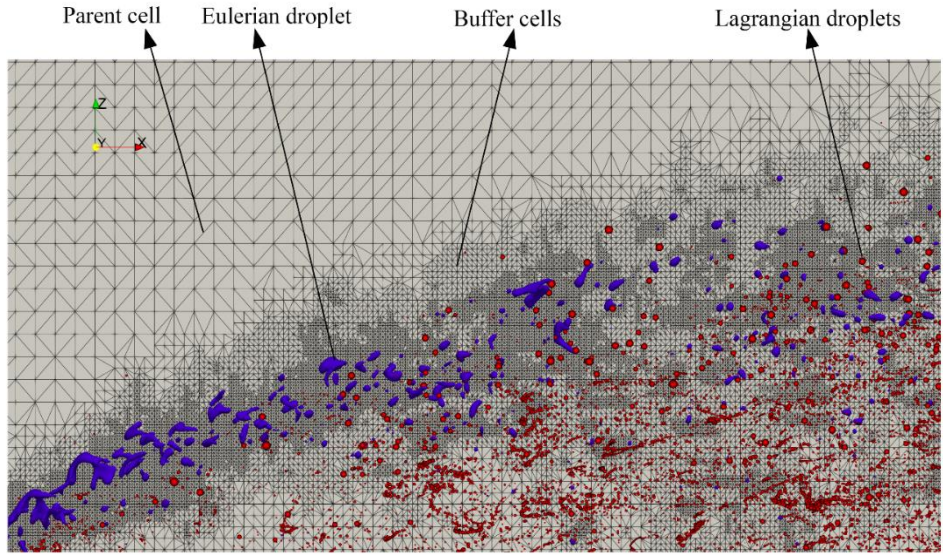

Fig. 4: Adaptive mesh refinement (AMR) of cells using liquid volume fraction alpha field. Red droplets indicate lagrangian droplets, which are converted from Eulerian to Lagrangian.

As the Eulerian liquid droplets move from one point to another, the cells where the interface is located are refined continuously during the simulation. Once a Lagrangian droplet forms (from an Eulerian framework, as shown in Figure 4), the grid is coarsened back to the parent base mesh resolution. In this way, the Lagrangian assumption also holds for droplets, as the liquid fraction is less than 0.22 in a computational cell (Arlov et al. (2007)). Note that Lagrangian droplets and Eulerian fields are shown in red and blue in Figure 4, respectively, are in 3-dimensions, whereas the AMR grid in the background is just a 2-dimensional, mid-section plane of the domain. Hence, many red (Lagrangian) droplets are in the fine mesh while the coarse grid surrounds Eulerian ones. Nearby cells in the vicinity of Eulerian droplets are also adjusted by adding six buffer layers to prevent excessive cell size jumps in the flow field, which could result in considerable pressure and velocity gradients at the jump point. The Eulerian and Lagrangian droplets in the computational domain, the AMR, and the buffer cells are shown in Figure 4. Depending on the extent of break-up and penetration of the liquid jet, the total number of cells in the domain increases from an initial count of 0.75 million to 3-5 million during the runtime. The maximum Courant Friedrich Lewy number (CFL) is set to 0.12, providing an average time step of $1.5 \times 10^{-8}$ sec.

The cross-flow air inlet is fed with a fully developed turbulent velocity profile obtained from a separate channel flow simulation. The liquid jet is provided a uniform velocity profile with negligible turbulence at the nozzle exit to avoid any early breakup due to turbulence (Li. et al., 2016). Water is used as the liquid, and the air is used as the crossflow fluid. The liquid jet and cross-flow air temperatures are kept at a constant room temperature of 25ºC for all test conditions. Table 1 and 2 describes the fluid properties and test conditions used for the simulations.

| Fluid Property | P=2.1 bar, T=25ºC | P=3.8 bar, T=25ºC |
|---|---|---|
| $\rho_{jet}$ (kg/m³) | 997.10 | 997.17 |
| $\rho_{air}$ (kg/m³) | 2.42 | 4.44 |
| $\sigma$ (N/m) | 0.072 | 0.072 |
| $\nu_{air}$ (m²/s) | 7.638 x 10⁻⁶ | 4.167 x 10⁻⁶ |
| $\nu_{jet}$ ( m²/s) | 8.927 x 10⁻⁷ | 8.925 x 10⁻⁷ |
| $\mu_{air}$ (N.s/m²) | 1.849 x 10⁻⁵ | 1.851 x 10⁻⁵ |
| $\mu_{jet}$ (N.s/m²) | 8.901 x 10⁻⁴ | 8.900 x 10⁻⁴ |

Table 1: Summary of fluid properties

| | Pressure (bar) | Crossflow velocity (m/s) | Jet velocity (m/s) |
|---|---|---|---|
| **Case 1** | 2.1 | 61 | 9 |
| **Case 2** | 2.1 | 61 | 12 |
| **Case 3** | 2.1 | 61 | 19 |
| **Case 4** | 2.1 | 61 | 24 |
| **Case 5** | 3.8 | 65 | 14 |
| **Case 6** | 3.8 | 65 | 19 |
| **Case 7** | 3.8 | 41 | 19 |
| **Case 8** | 3.8 | 41 | 9 |
| **Case 9** | 3.8 | 41 | 24 |
| **Case 10** | 3.8 | 33 | 19 |

Table 2: Simulated test conditions with nozzle diameters of 572 µm.

# 3. Results and Discussion

This section first assesses our VOF-LPT coupled framework by performing a validation case at elevated pressures of 2.1 and 3.8 bars, which would mimic the density ratios of actual gas turbines. A grid test is then carried out to investigate the effect of grid sizes on droplet sizes and jet penetration. And the following section discusses the impact of various parameters on droplet size characteristics and compares our jet trajectory against various experimental correlations. Further, we discuss the primary break-up behavior and flow features (vortex formations) observed in the liquid jet in crossflow. All the determined parameters ($D_{32}$, STD) are time-averaged values calculated within the sub-domain (blue-colored), as shown in Figure 2.

## 3.1 Spray Trajectory and Validation

The compressible VOF-LPT model is validated using a Liquid Jet-in Crossflow (JICF) case at elevated pressure and room temperature conditions. The numerical results of the windward trajectory, droplet sizes ($D_{32}$), and the standard deviation (STD) are compared against the experimental results from Amighi and Ashgriz (2019), as shown in Figure 5. The validation is performed for two cases with nozzle diameters of 572 µm (Case A) and 457 µm (Case B). The parameters used with cases A and B are listed in Table 3.

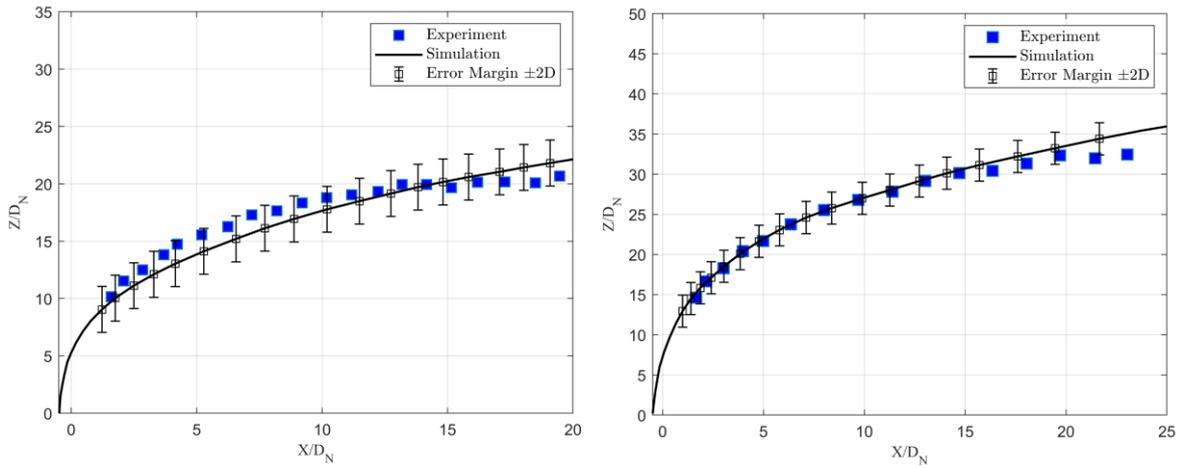

Fig. 5: Comparison of liquid jet trajectory vs. experimental for cases A and B. The error margins of $\pm 2d$ are considered based on the simulations due to the lack of error margin in experiments.

|  | **Case A** | **Case B** |
|---|---|---|
| Pressure | 3.8 bar | 2.1 bar |
| Temperature | 25 ℃ | 25 ℃ |
| Liquid jet velocity | 19 m/s | 19 m/s |
| Crossflow velocity | 65 m/s | 50 m/s |
| Nozzle diameter | 572 µm | 457 µm |

Table 3: Parameters used for numerical simulations for case A and case B

As shown in Figure 5, the error bars represent $\pm 2d$ the margin for both cases A and B. In this work, the error bars are considered based on the computational data because of the lack of data on error margins in experimental results. All the trajectory calculations are employed by assuming the center of the nozzle as the origin (0,0). As observed in Figure 5, the windward trajectory starts from -0.5D with respect to the origin on the x-axis. For the low-pressure case of 457 µm as nozzle diameter (case B), the predicted trajectory is closer in the near nozzle region (<20D) and deviates slightly after 20D. The deviation observed in the region >15D for case A and >20D for case B is because the experimental trajectory data is based on the spray plume. An image averaging and thresholding technique is used to determine the windward trajectory. This is chosen to account for the Eulerian and Lagrangian droplets' contributions to the trajectory. For the Eulerian contribution to a trajectory, an iso-contour value is used. The droplets are scaled to exact sizes for the lagrangian part to generate the whole spray atomization image. For cases A and B, the trajectory is plotted by averaging liquid spray images over time, where approximately 200 images are obtained with a time interval of 0.05 milliseconds between each image. A threshold of 0.9 is applied to the averaged image, which is sufficient to remove the traces of stray droplets outside the windward side of the trajectory (Gopala et al., 2010).

|  | Case A | | Case B | |
|---|---|---|---|---|
|  | $D_{32}$ (μm) | STD (μm ) | $D_{32}$ (μm) | STD (μm ) |
| Experimental | 74.5 | 18.0 | 84.5 | 22.1 |
| Computational | 71.96 | 19.9 | 78.13 | 20.71 |
| Error % | 3.4 | 10.5 | 7.5 | 6.3 |

Table 4: Comparison of Sauter Mean Diameter (D32) and the Standard Deviation (STD) for computational and experimental cases.

Table 3 presents the droplet sizes obtained numerically and experimentally for cases A and B. The $D_{32}$ values obtained for the validation cases A and B lie within a 10% error margin concerning the experimental $D_{32}$. Similarly, the error observed in the STD is also within a 12% error margin. Regarding the error margin observed here, the numerically predicted droplet sizes and standard deviation are much closer to the experimental results. This reveals a good agreement of the predicted data against experiments. Thus, comparing the spray trajectory with Sauter Mean Diameter ($D_{32}$) and Standard Deviation (STD) of droplets against the experiments illustrates the accuracy of the numerical simulation.

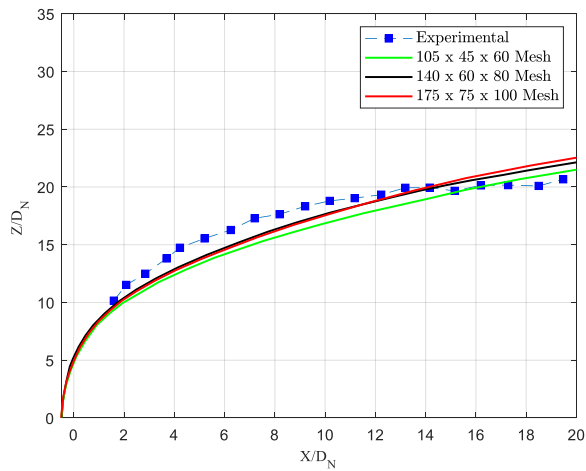

Fig. 6: Comparison of trajectory for different mesh resolutions with experimental

Three different baseline meshes are chosen to study grid independence, a coarse grid (105 x 45 x 60), a medium grid (140 x 60 x 80), and a fine grid (175 x 75 x 100). The windward trajectory and the spray droplet size characteristics, such as the Sauter mean diameter (SMD or $D_{32}$) and standard deviation (STD) for these grids, are compared in Figure 6 and Table 5. For the near nozzle region (<20D) considered here, the variation in trajectories is minimal, and there isn't much difference between the trajectories. However, for the medium and fine grids, the trajectory is traced along the same path until 10D, and later on, only a slight difference is observed. Compared to the medium and fine grids, the trajectory is slightly underpredicted for the coarser grid. As a result, we can infer that the grid has less impact on the trajectory near the nozzle (<20D) and that the trajectory is closer to experimental data in this area. This could be attributed to the adaptive mesh refinement (AMR) taking care of the refinement to some extent.

| | $D_{32}$ (μm) | STD (μm) | Error in $D_{32}$ | Error in STD | Initial cell count (x $10^6$) | Final cell count (x $10^6$) |
|---|---|---|---|---|---|---|
| **Experimental** | 74.5 | 18.0 | - | - | - | - |
| **Coarse** | 78.75 | 21.62 | 5.7 % | 20.1 % | 0.35 | 1.27 |
| **Medium** | 71.96 | 19.9 | 3.4 % | 10.5 % | 0.75 | 3.2 |
| **Fine** | 71.07 | 20.08 | 4.6 % | 11.5 % | 1.41 | 6.75 |

Table 5: Sauter Mean Diameter ($D_{32}$) and the Standard Deviation (STD) and error observed for different mesh resolutions

For the medium and fine grids, the values of SMD and STD are found closer to each other and close to experimental values. The SMD and STD values show significant overprediction for the coarser grid. The error observed in STD for the coarser grid is twice that of medium and fine grids. Therefore, our medium-sized grid (140 x 60 x 80) can provide sufficiently accurate and computationally less expensive results for droplet sizes and liquid jet penetration. This medium grid will be used in all our analyses from this point onwards. It is observed that when AMR is used, the final cell count is approximately four-five times the initial cell count in the domain. The total cell count observed for different simulation test cases varies between 3-5 million grid cells depending on the liquid jet penetration and break-up behavior.

## 3.2 Effect of various parameters on droplet size characteristics

This section talks about the effect of various parameters, the liquid jet velocity ($V_j$), cross-flow velocity ($V_{air}$), and ambient pressure (P), on the droplet size characteristics, namely the Sauter mean diameter ($D_{32}$) and the standard deviation (STD). All the test cases are carried out at constant temperatures of 25°C and high-pressure conditions of 2.1 and 3.8 bar. Table 6 summarises all the droplet size characteristics gathered from simulations and experiments. The momentum flux ratio and crossflow

Weber numbers of the test cases range from 8 to 80 and 38 to 150, respectively. The number of lagrangian droplets formed ranges from 40,000 to 3,00,000 depending on the extent of penetration and break-up for the test cases performed.

| | | Momentum flux ratio (q) | Weber number (air) (We) | Experimental | | Computational | | Error | |
|---|---|---|---|---|---|---|---|---|---|
| | | | | $D_{32}$ (µm) | STD (µm) | $D_{32}$ (µm) | STD (µm) | $D_{32}$ (%) | STD (%) |
| **Case 1** | P=2.1bar,T=25°C, Va=61m/s, Vj=9m/s | 8 | 71 | 88.8 | 23.9 | 81.93 | 22.77 | 7.74 | 5.02 |
| **Case 2** | P=2.1bar,T=25°C, Va=61m/s, Vj=12m/s | 16 | 71 | 84.3 | 22.1 | 79.16 | 21.6 | 6.09 | 2.26 |
| **Case 3** | P=2.1bar,T=25°C, Va=61m/s, Vj=19m/s | 41 | 71 | 78.5 | 19.6 | 76.15 | 20.89 | 2.99 | 6.58 |
| **Case 4** | P=2.1bar,T=25°C, Va=61m/s, Vj=24m/s | 66 | 71 | 73.9 | 17.6 | 72.24 | 19.21 | 2.24 | 9.15 |
| **Case 5** | P=3.8bar,T=25°C, Va=65m/s, Vj=14m/s | 10 | 150 | 78.8 | 19.8 | 75.49 | 21.98 | 4.20 | 11.01 |
| **Case 6** | P=3.8bar,T=25°C, Va=65m/s, Vj=19m/s | 19 | 150 | 74.5 | 18.0 | 71.96 | 19.9 | 3.4 | 10.5 |
| **Case 7** | P=3.8bar,T=25°C, Va=41m/s, Vj=19m/s | 48 | 60 | 79.3 | 20.0 | 75.48 | 20.43 | 4.81 | 2.15 |
| **Case 8** | P=3.8bar,T=25°C, Va=41m/s, Vj=9m/s | 10 | 60 | 87.6 | 23.4 | 82.95 | 22.56 | 5.31 | 3.59 |
| **Case 9** | P=3.8bar,T=25°C, Va=41m/s, Vj=24m/s | 77 | 60 | 75.2 | 18.1 | 68.83 | 17.55 | 8.47 | 3.03 |
| **Case 10** | P=3.8bar,T=25°C, Va=33m/s, Vj=19m/s | 77 | 38 | 78.6 | 19.6 | 78.15 | 21.2 | 0.57 | 8.16 |

Table 6: Summary of test results: Sauter Mean Diameter ($D_{32}$) and the Standard deviation for all test cases ($D_N$=572 µm)

For all the test cases performed, the maximum error observed on D32 and STD is 8.47 % and 11.01 %, respectively, and the average error observed on D32 and STD are 4.58 % and 6.15 %, respectively. For all the D32 plots (Figures 7, 8, and 9), an error margin of 10 % is provided on the computational D32 plot, which corresponds to an average error of $\pm 7.5 \mu m$. Similarly, an error margin of 15% is provided on the numerically calculated STD. All the D32 values obtained are observed to be within this range. Since the experimental error values are unknown, the error margin is provided concerning the numerical data. The values of D32 are found to be on the lower side for all the test cases performed compared to the experimental observations. D32 and STD values are calculated in every simulation by iterating over all the Lagrangian particles throughout the domain. To maintain the accuracy of results similar to the experimental procedure, the irregular Eulerian droplets are neglected from the calculations, including the ligaments carried out in the experiments. The sphericity and the minimum threshold value of the droplet undergoing conversion from one framework to another are set per the experimental data analysis. This way, it ensures the droplets converted to Lagrangian particles are the ones that need to be accounted for in the droplet characteristics calculations. Each data point on the plot represents either simulation or experimental data. The following sections discuss the effect of liquid jet velocity, crossflow velocity, and ambient pressure on the droplet size characteristics.

### 3.2.1 Effect of Liquid Jet Velocity/ Momentum flux ratio

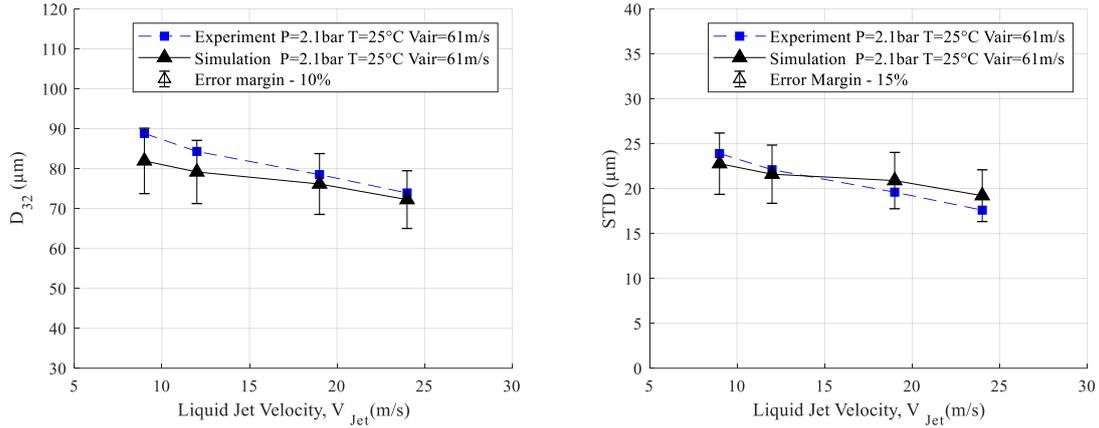

Fig. 7: Plot of Sauter Mean Diameter (D32) and Standard Deviation (STD) with Liquid jet velocity at a pressure of 2.1bar. The 10% and 15% error margins are considered for D32 and STD, respectively, based on the simulations due to the paucity of error margin in experiments.

In all of the $D_{32}$ plots, the error bars correspond to an error margin of 10% applied on the simulated contour, and on STD plots, it corresponds to 15 %. Typically, the errors in measurements involving particle statistics (e.g. $D_{32}$, STD) vary between 10-30%. Considering this, our predictions exhibit an excellent agreement regarding the accuracy of the droplet statistics. Figure 7 shows the effect of jet velocity on global droplet sizes ($D_{32}$ and STD) at an ambient pressure of 2.1 bar. From both the numerical and experimental observations, the effect of liquid jet velocity is to decrease the size of the droplets (both $D_{32}$ and STD) formed. This is attributed to the increased atomization from the higher momentum flux ratio (q). Momentum flux ratio (q) is defined as:

$$q = \frac{\rho_j V_j^{\,2}}{\rho_{air} U_{air}^{\,2}} = \frac{We_{jet}}{We_{air}}$$

(29)

The droplet sizes decrease as the jet velocity increases from 9 m/s to 24 m/s. Similarly, for the standard deviation, it is also reduced. The lower standard deviation due to increased jet velocity indicates that the atomization is more uniform at higher jet velocities. The reduction in droplet size is attributed to mainly two factors. (1) One is that the Reynolds number increases, and the jet becomes more turbulent on increasing the jet velocity, resulting in an increased break-up. (2) Another factor is that as the jet velocity increases, the jet is penetrated more into the cross-flow. This increases the exposure of the liquid jet with the cross-flow. These two factors combined result in the decrease of $D_{32}$ and STD values. The above plot shows that increased liquid jet velocity produces finer droplets, improving the atomization process.

### 3.2.2 Effect of Cross-flow Velocity

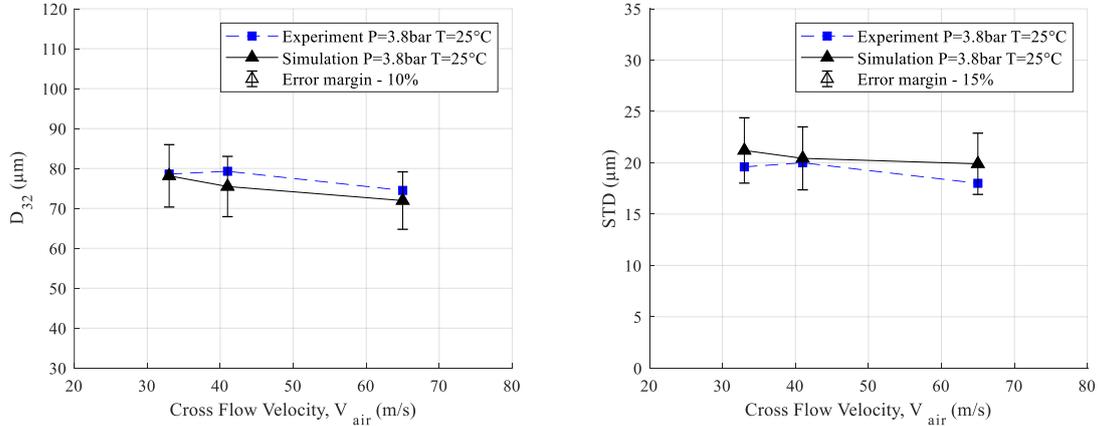

Fig. 8: Plot of Sauter Mean Diameter (D32) and Standard Deviation (STD) with crossflow velocity at a pressure of 3.8 bar. The error margins of 10% and 15% are considered for $D_{32}$ and STD, respectively.

Figure 8 shows the effect of cross-flow velocity on $D_{32}$. The cross-flow velocity is varied from 33 m/s to 41m/s and further to 65 m/s, where all other parameters, the pressure (P), liquid jet velocity ($V_j$), temperature (T), are kept constant. As observed above, the increase in cross-flow velocity decreases the droplet size. At low cross-flow velocity, the penetration of the liquid jet is higher. The increased penetration is due to the increased momentum flux ratio from the reduced air velocity. The penetration is higher for lower air velocity because the drag force exerted on the liquid jet is smaller at lower cross-flow velocity. This reduces the width of the spray plume, resulting in the decreased interaction between the liquid and the gas. On the other hand, at higher cross-flow velocity, the higher drag force bends and flattens the liquid jet further, resulting in a more pronounced break-up and atomization. Another explanation for the reduction in droplet sizes is that as the crossflow velocity increases, the Weber number of the air increases. This also causes the break-up mode to be shifted to pure shear mode (We>110), resulting in the production of a large number of smaller droplets, also causing a reduction in droplet sizes. Weber number of air:

$$We = \frac{\rho_a U_a{}^2 D_N}{\sigma} \qquad (30)$$

The experimental $D_{32}$ value at a crossflow velocity of 33m/s is slightly lower than the value at 41m/s. But our computational prediction in Figure 8 shows that the $D_{32}$ value increases with a decrease in the crossflow velocity. The slight discrepancy in the experimental data could be due to any experimental errors, and this particular trend wasn't observed with other experimental results pertaining to different crossflow velocities. The general trend for the droplet sizes is to decrease with increased crossflow velocity. Therefore, the numerically predicted trend is more acceptable here.

### 3.2.3 Effect of Pressure

Figure 9 shows the effect of cross-flow pressure on Sauter mean diameter. The ambient pressure varies while the cross-flow velocity (61 & 65 are considered the same) and liquid jet velocity are kept constant. As seen above, increasing the pressure decreases the SMD. A similar observation is also obtained in numerical simulation. This is because an increase in pressure increases density and the drag force on the jet and the ligaments and droplets, breaking into smaller droplets. The value of the Sauter mean diameter depends more on the larger droplets than the smaller droplets. So a reduction in the generation of larger droplets consequently reduced the SMD. The STD plot for increased pressure also shows similar behavior. The increased pressure improves atomization, resulting in a more uniform droplet size. In addition, the enhancement in the pressure decreases the penetration of the jet due to the increased drag forces on the liquid jet.

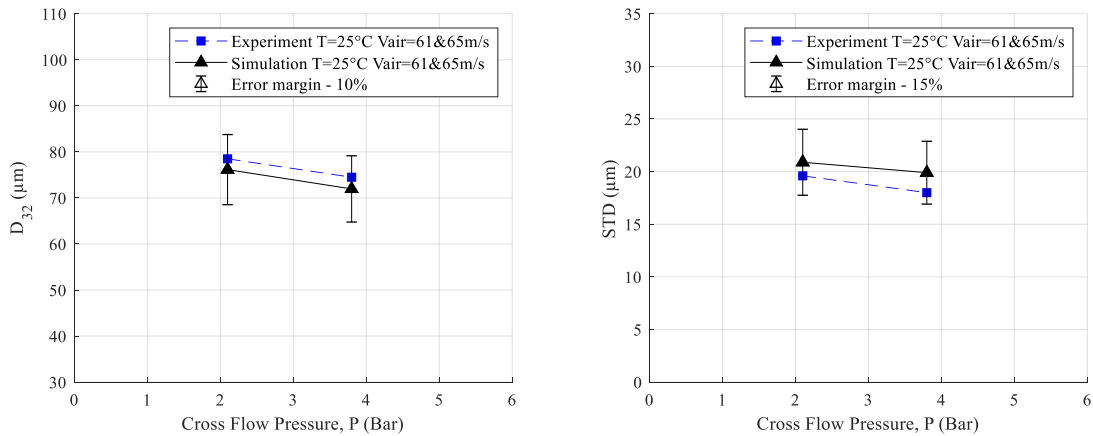

Fig. 9: Plot of Sauter Mean Diameter ($D_{32}$) and Standard Deviation (STD) with ambient pressure at a crossflow velocity of 65 and 61 m/s. The error margins of 10% and 15% are considered for D32 and STD, respectively.

### 3.2.4 Effect of Cross-flow Weber number

Here we have considered two sets of cases (case 5, 8 and case 9, 10) having the same liquid-to-gas momentum flux ratio (q) with different cross-flow Weber numbers (We). The cross-flow and liquid jet velocities are varied proportionately to keep the momentum flux ratio constant (q=10 for cases 5, 8 and q=77 for cases 9, 10). The momentum flux ratio is:

$$q = \frac{\rho_j V_j^2}{\rho_{air} U_{air}^2} = \frac{We_{jet}}{We_{air}}$$

(31)

As the Weber number is increased (Case 8: q=10, We=60 and Case 5: q=10, We=150) from 60 to 150, the droplet sizes are reduced (Case 8: $D_{32}$=82.95, STD=22.56 and Case 5: $D_{32}$=75.49, STD=21.98) following a general trend (Lubarsky et al.,2010; Lee et al., 2007). The same observation is also made at a higher momentum flux ratio (q=77) when the Weber number is increased from 38 to 60 (Case 10: $D_{32}$=78.15, STD=21.20 and Case 9: $D_{32}$=68.83, STD=17.55).

### 3.2.5 Effect of parameters on liquid jet penetration, breakup region, and droplet distribution
### (a) Liquid Jet Penetration – trajectory

Many researchers have proposed empirical correlations in power law, logarithmic, exponential, etc., to predict penetration heights for a liquid jet in crossflows. Some researchers have considered the effect of momentum flux ratio alone (Tambe et al., 2005), while several others included the effect of Weber number, pressure, viscosity, etc., in their correlations (Amighi and Ashgriz (2019), Ragucci et al. (2007), and Elshamy et al. (2007)). Stenzler et al. (2006) proposed a power-law correlation that accounts for fluid viscosity and aerodynamic weber number and is among the first to assess the effect of air viscosity on jet penetration. While Becker and Hassa (2002), Chen et al. (1993) pointed out that there is no significant effect of aerodynamic weber number and break-up mode on liquid jet penetration. The correlations proposed by Amighi and Ashgriz (2019), Tambe et al. (2005), Ragucci et al. (2007), and Elshamy et al. (2007) for the windward trajectories are used to compare against our numerically obtained trajectories. Most of the available correlations in the literature are only valid in the near nozzle region (<25D).

Tambe : $\quad \dfrac{y}{D_N} = 1.55 q^{0.53} \ln\left(1 + 1.66\dfrac{z}{D_N}\right)$

(32)

Ragucci : $\quad \dfrac{z}{D_N} = 2.27 q^{0.44} We_{aero}^{-0.012}\left(\dfrac{x}{D_N}\right)^{0.367}$ $\quad$ and $\quad \dfrac{z}{D_N} = 2.28 q^{0.422} We_{aero}^{-0.015}\left(\dfrac{\mu}{\mu_{air,300k}}\right)^{0.186}\left(\dfrac{x}{D_N}\right)^{0.367}$

(33)

Elshamy : $\quad \dfrac{y}{D_N} = 12.63 q^{0.446}\left[1 - e^{-\left(\frac{x}{D_N}+0.5\right)/10.46}\right] \times \left[1 + 1.42 e^{-\left(\frac{x}{D_N}+0.5\right)/4.14}\right] \times \left[1 + e^{-\left(\frac{x}{D_N}+0.5\right)/1.39}\right] \times We^{-0.141}\left(\dfrac{p}{p_o}\right)^{-0.05}$

(34)

Amighi : $\quad \dfrac{y}{D_N} = 13.31\left(\dfrac{x}{D_N}+0.5\right)^{0.27} q^{0.40} We_{air}^{0.14} Oh_{air}^{-0.20} Oh_{Jet}^{0.65}$

(35a)

$\quad = 13.31\left(\dfrac{x}{D_N}+0.5\right)^{0.27} We_{air}^{-0.35} We_{jet}^{0.72} Re_{air}^{0.20} Re_{Jet}^{-0.65}$

(35b)

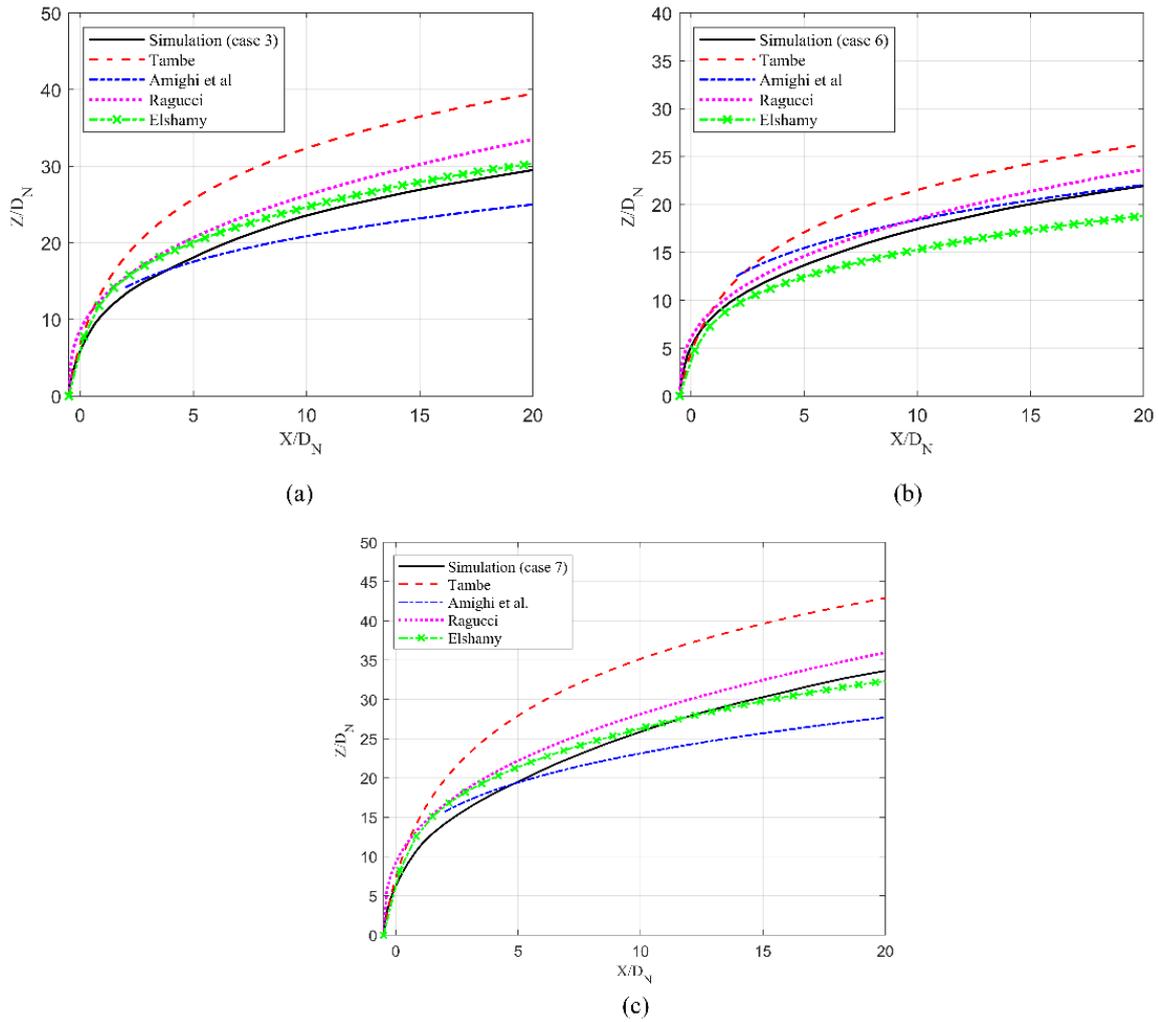

Fig. 10. Liquid jet penetration along with correlations for (a) case 3 (P=2.1 bar), (b) case 6 (P=3.8 bar) and, (c) case 7 (P=3.8 bar)

In Figure 10, the numerically obtained trajectories at two different pressure conditions (2.1 and 3.8 bar) are compared against the experimental correlations of Amighi, Tambe, Ragucci, and Elshamy. In Figure 10(a), the trajectory for case 3 with pressure 2.1 bar agrees well with the Elshamy correlation, while the Amighi correlation under-predicts it. Figures 10(a) and 10(b) have similar crossflow velocities of 61 m/s and 65 m/s, respectively, and the same liquid jet velocity of 19 m/s. The pressure change has a consequent effect on the Weber number and momentum flux ratio because of the change in crossflow air density. The weber number reduces to half while the momentum flux ratio doubles. The Elshamy correlation underpredicts the trajectory in Figure 10(b), whereas Amighi's correlation closely predicts in this case. Amighi's inverse dependence of windward side trajectory on the weber number is why it shows minor change compared to other correlations

All correlations show similar predictions in Figure 10(a) and Figure 10(c) regarding the trajectory of the current simulation. The weber number and momentum flux ratio are almost identical for cases 3 and 7. Therefore, the simulated data lies between the Elshamy and Amighi's correlation, and the minimal change is only caused by the slight differences in the momentum flux ratio and weber number values. Hence, the effect of change in pressure is sufficiently reflected through these non-dimensionless numbers. The correlation of Tambe highly over-predicts the trajectory in the analysis of all the cases, which can be attributed to the absence of the weber number in empirical correlation.

## (b) Liquid jet break-up region

| | | Momentum flux ratio (q) | Weber number (We) (air) | Breakup Location | |
| --- | --- | --- | --- | --- | --- |
| | | | | X (in $D_N$) | Y (in $D_N$) |
| **Case 1** | P=2.1bar,T=25°C, Va=61m/s, Vj=9m/s | 8 | 71 | 8.62 ± 1.37 $D_N$ | 9.95 ± 0.484 $D_N$ |
| **Case 2** | P=2.1bar,T=25°C, Va=61m/s, Vj=12m/s | 16 | 71 | 8.785 ± 2.33 $D_N$ | 14.161 ± 0.964 $D_N$ |
| **Case 3** | P=2.1bar,T=25°C, Va=61m/s, Vj=19m/s | 41 | 71 | 9.64 ± 1.73 $D_N$ | 22.69 ± 1.061 $D_N$ |
| **Case 4** | P=2.1bar,T=25°C, Va=61m/s, Vj=24m/s | 66 | 71 | 9.29 ± 1.385 $D_N$ | 27.84 ± 1.23 $D_N$ |
| **Case 5** | P=3.8bar,T=25°C, Va=65m/s, Vj=14m/s | 10 | 150 | 8.79 ± 0.827 $D_N$ | 11.5 ± 0.59 $D_N$ |
| **Case 6 (Case A)** | P=3.8bar,T=25°C, Va=65m/s, Vj=19m/s | 19 | 150 | 8.67 ± 0.873 $D_N$ | 15.335 ± 0.622 $D_N$ |
| **Case 7** | P=3.8bar,T=25°C, Va=41m/s, Vj=19m/s | 48 | 60 | 11.3 ± 0.933 $D_N$ | 26.14 ± 0.51 $D_N$ |
| **Case 8** | P=3.8bar,T=25°C, Va=41m/s, Vj=9m/s | 10 | 60 | 8.85 ± 1.17 $D_N$ | 11.11 ± 0.73 $D_N$ |
| **Case 9** | P=3.8bar,T=25°C, Va=41m/s, Vj=24m/s | 77 | 60 | 8.40 ± 1.08 $D_N$ | 30.10 ± 1.58 $D_N$ |
| **Case 10** | P=3.8bar,T=25°C, Va=33m/s, Vj=19m/s | 77 | 38 | 10.22 ± 1.12 $D_N$ | 31.58 ± 1.83 $D_N$ |

Table 7. The streamwise and transverse location of breakup region for cases 1 to 10.

The break-up location is investigated for all the cases (Table 6), and it is observed that the break-up does not occur precisely at a particular point. Instead, the break-up occurs in a broader region in the streamwise direction than in the transverse direction (jet direction). Therefore, we have considered break-up location a region rather than a particular point where the break-up occurs. The break-up location refers to the region where the liquid core, after bending, shows excessive deformation and discontinuities in the liquid core of the spray trajectory. This is the same as the column break-up location, except our break-up is more in a multimode/shear break-up mode where both the bag-shear and shear break-up are observed. The liquid jet is subjected to unsteady aerodynamic forces, causing the windward and leeward surfaces to fluctuate, resulting in the liquid jet column's deformation, bending, and fracture. The break-up location is determined for several test cases, and it is found that the x-location of the break-up is almost constant. As shown in Table 7, it does not show significant variation in the streamwise direction. It is approximately located at 9.2D±1.2D for all the cases, whereas the y-location seems to vary by a large magnitude. The y-location varies with the jet penetration, dependent on the momentum flux ratio. The higher the momentum flux ratio, the more the y-location of the jet breakup. Wu et al. (1997) also investigated the break-up location and found the column fracture location constant at about 8D downstream of the nozzle. Compared to the break-up location by Wu et al. (1997), our break-up is slightly delayed and in good agreement with the experiments. Tambe et al. (2005) also made a similar observation regarding the independence of momentum flux ratio on streamwise break-up location. While the transverse location of break-up varies with the crossflow parameters (momentum flux ratio and crossflow weber number), the momentum flux ratio has a more significant influence on penetration than the crossflow weber number.

### (c) Droplet Distribution

In all of the test cases performed, the larger droplets or the liquid lumps are found closer to the upper periphery of the liquid jet, i.e. in the upper half of the spray core region. These larger, irregular fragments are formed from the liquid core break-up and are penetrated more into the crossflow due to the higher momentum of the larger droplets, which are less affected by the crossflow. The smaller droplets are larger in number and found more in the lower part. The smaller droplets are produced in two ways: one from the secondary breakup of larger droplets and another due to the shear break-up from the sides of the liquid column. At higher crossflow weber numbers, this model of a shear break-up from the sides of the liquid column is more dominant than the other break-up modes leading to the production of a large number of smaller droplets causing a decrease in the droplet sizes. For higher momentum flux ratio cases (Case 4,7,9,10 – Table 6), the droplet sizes are found to peak at the upper periphery of the spray core, while for lower momentum flux ratio cases (Case 1,5,8), the droplet sizes peak near the spray core but still lies in the upper half region.

Figure 11 shows the Eulerian and Lagrangian droplets produced in crossflow atomization. The droplets subjected to more deformation stay in the Eulerian field, and those earlier converted from Eulerian to Lagrangian are visible in the zoomed-in image. This conversion is facilitated by the droplet sphericity, which measures the deviation in droplet shape compared to that of a spherical droplet of equivalent diameter. The Eulerian droplets visible in Figure 11 are the droplets having sphericity >1.47, and the spherical droplets visible are Lagrangian droplets. The simulation produces as small as 10-15 µm droplets, better resolved using the LPT approach. The larger droplets are fewer in number, and the smaller droplets are larger in number. The droplet sizes generally follow a log-normal distribution (Li et al. (2016)).

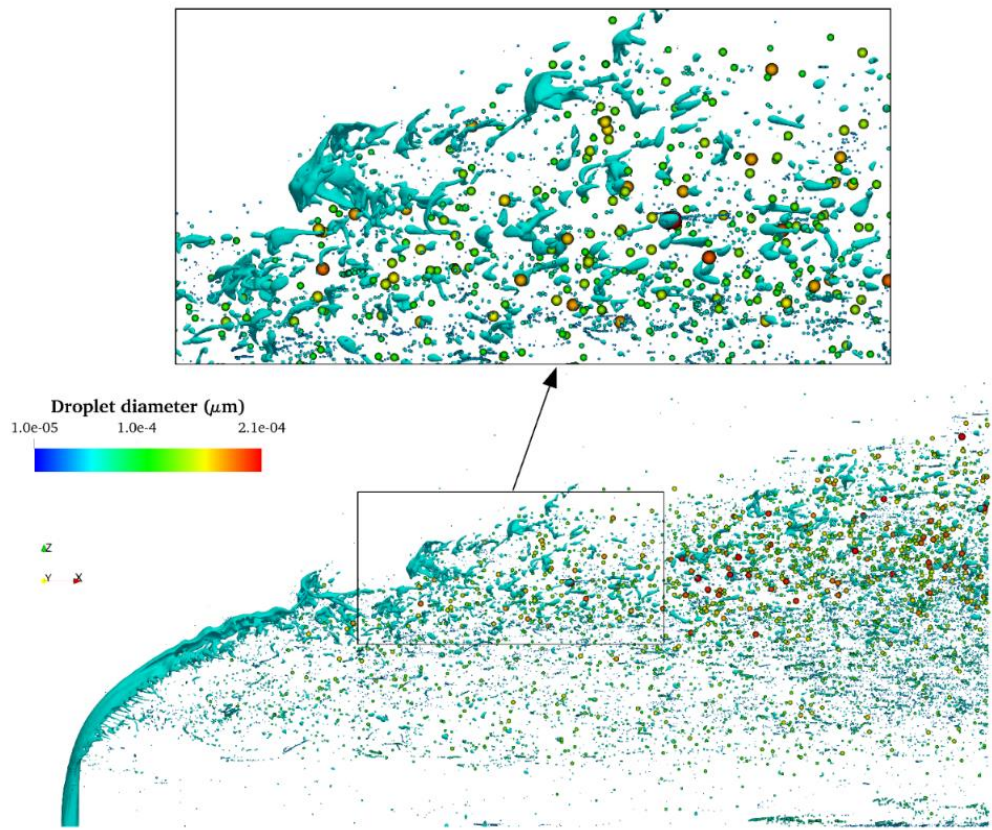

Fig. 11: Crossflow jet atomization and breakup captured using VOF-Lagrangian particle tracking approach for case 5 (P=3.8bar, Vair=65m/s, Vj=14m/s, q= 10, We=150). Spherical droplets indicate lagrangian droplets converted from Eulerian droplets (blue color)

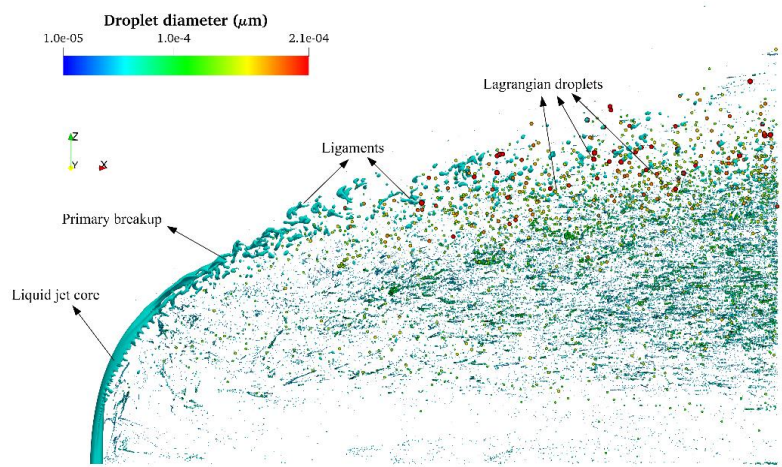

Fig. 12: Crossflow jet atomization and breakup captured using VOF-Lagrangian particle tracking approach for case 6 (P=3.8bar, Vair=65m/s, Vj=19m/s, q= 19, We=150)

## 3.3 Spray breakup and atomization

Figure 12 represents a liquid jet's break-up and atomization process in cross-flow (case 6) with Eulerian (iso-contour α=0.5) and the Lagrangian droplets. The aerodynamic drag force experienced by the liquid jet column forces it to bend in the crossflow direction because of a high-pressure windward and low-pressure leeward region, as shown for case 3 (Table 6) in Figure 13(a). Another prominent feature of LJICF is flattening the liquid jet cross-section from a circle to a crescent. This can be understood as airflow around a deformable cylinder, as shown in Figure 13(b). The crucial factors responsible for this deformation are – internal boundary layer flow and pressure difference. The internal boundary layer (shown by the dashed line in the first image of Figure 13(b)) is formed in the liquid phase due to the shear generated by the external flow of air around the liquid core, which transports the liquid from point 2 towards point 3 and 4. Moreover, the pressure at point 2 is higher than the pressure at points 3 and 4; it further helps this internal flow of the liquid away from the frontal region (around point 2) towards the periphery (near points 3 and 4). It leads to the flattening of the liquid jet core cross-section (as shown in the second image of Figure 13(b)). Both the factors continue forcing the liquid from point 2' towards 3' till the complete circular cross-section has been deformed into a thin sheet-like cross-section (third image of Figure 13(b)). The liquid movement from points 2'' to 3'' (in the third image of Figure 13(b)) leads to the thickening the liquid-jet core edges. This process is suspected to be one of the deciding factors in streamer formation along the two edges of the deformed liquid jet, which will be discussed later in detail. However, the flattening of the liquid jet core further increases the effective frontal area, and consequently, the drag forces cause it to deflect even further. Figure 13(c) shows this deformation of the liquid column for case 3 at varying distances from the bottom wall.

The liquid jet deformation is mainly followed by break-up through two mechanisms: column break-up and surface break-up. The column break-up is characterized by instabilities and the growth of surface waves on the liquid jet column, resulting in the formation of crests and troughs. These waves grow in size, causing the liquid column to detach from one of the wave's troughs, resulting in a liquid break-up. Large fragments separate in this break-up, leading to larger droplets observed in the liquid jet's upper periphery. In the surface break-up, the droplets are pinched off (before the column break-up occurs) from the sides of the liquid column due to the action of shear between the liquid and gas phases on the periphery of the flattened/deformed liquid jet as observed at height (Z) between 8D to 12D in Figure 13(c).

It is also proposed that the surface break-up is caused by the growth of turbulent instability on the liquid column as the laminar liquid jet undergoes transition (Madabhushi et al., 2008). This is commonly observed at moderate to high crossflow Weber numbers ($We_{air}$) with high momentum flux ratios (q). At lower to moderate Weber numbers, increasing crossflow weber number shifts the break-up towards the shear regime of a column breakup. The shear break-up starts way before the instabilities at the liquid-gas interface, causing the liquid core to rupture into smaller droplets. The droplets formed from shearing action are smaller, resulting in better atomization of the liquid jet. These two break-up processes constitute the primary atomization part.

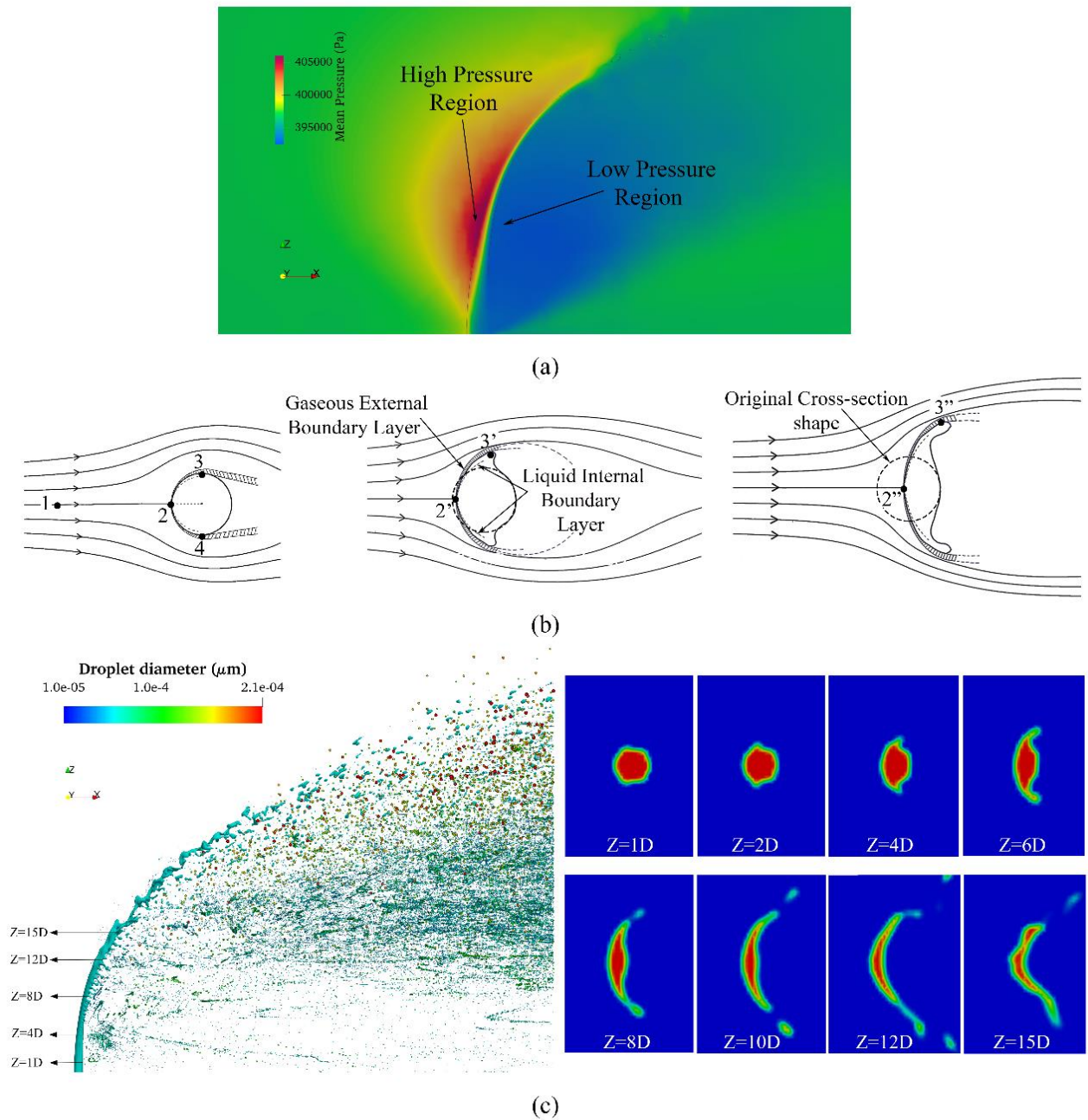

Fig. 13: (a) Mean pressure showing maximum and minimum values on the windward and leeward side of a liquid column, (b) Schematics showing liquid column deformation at increasing distance (height) from the point of injection, (c) Liquid jet cross-section deformation along with shear breakup at various heights from the point of injection (in jet direction, Z) for case 3 (P=2.1bar, $V_{jet}$=19m/s, q=41, $We_{air}$=71).

The primary break-up happens when the aerodynamic forces of the crossflow cause the liquid jet to rupture into smaller ligaments and droplets. Sallam et al. (2004; 2006) classified the primary break-up for non-turbulent liquid jets based on the aerodynamic Weber number into four modes, the column (We<4), bag (4<We<30), multimode (30<We<110) and shear breakup modes (We>110). The break-up mode is investigated for different crossflow weber numbers (38-150) and momentum flux ratios (8-77). For the test cases performed in this study, the breakup modes involve bag/multi-mode and shear modes of breakup and the surface breakup of the jet. For a few cases, multimode behavior is observed where both bag and shear modes are present. As the aerodynamic weber number increases, the mode of break-up shifts from multimode to pure shear mode (cases 5 and 6), characterized by the pinch-off of droplets from the sides of the liquid column. The bag formation and its subsequent breakage into ligaments and droplets for case 8 are discussed in detail in the subsequent section. Similar characteristics of multimode and shear breakup modes are discussed later in the following sections.

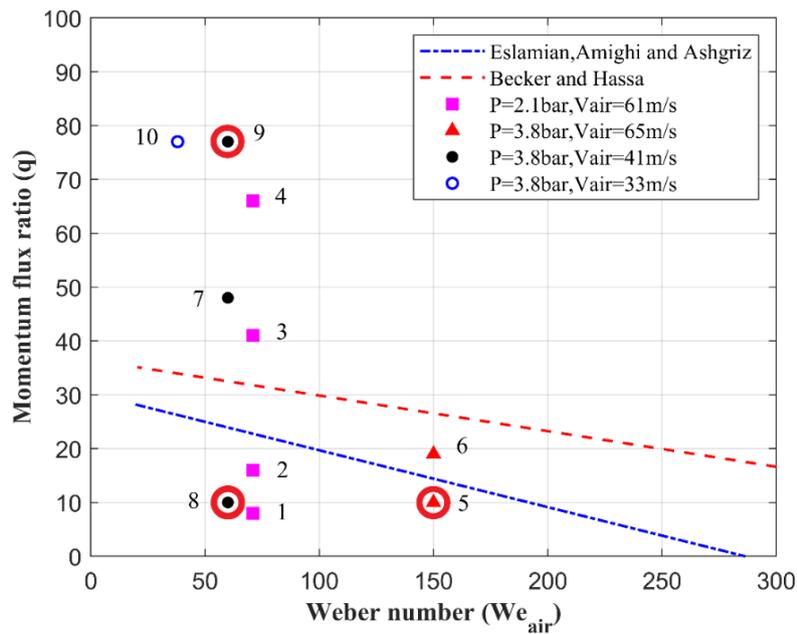

Fig. 14. The primary breakup regime map shows the varying momentum flux ratios and crossflow Weber numbers cases. The red-colored encircled points show the cases that are analyzed in detail.

Three tests out of ten listed in Table 6 are considered for detailed analysis. The three tests belong to the different regimes of the liquid jet breakup, as shown in Figure. 14 – case 8 in bag breakup regime of the column breakup, case 9 in surface breakup regime, and case 5 in shear breakup regime of the column breakup.

### 3.3.1 Bag Breakup Regime

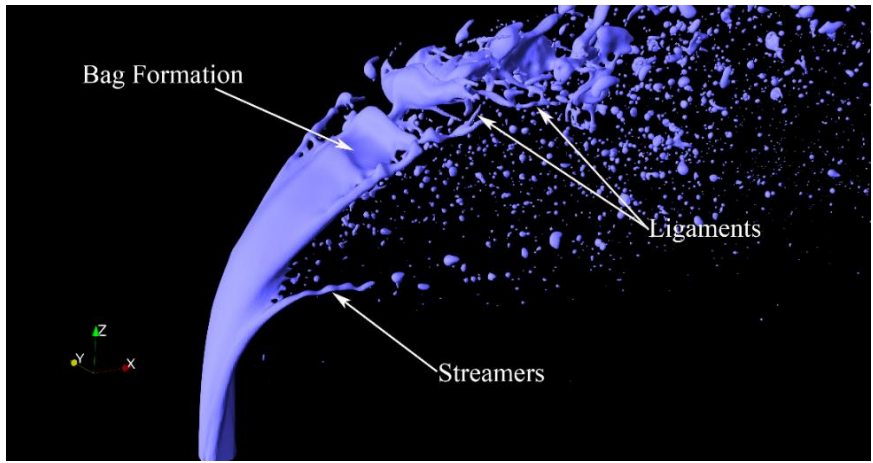

Fig. 15. Characteristic features of a LJICF in bag breakup regime for case 8 (q = 10 and We = 60).

Case 8 is ideal for bag breakup because of the low-value q and We. A few distinct features of this case are the flow in the wake of the liquid jet, development of instability and its growth, bag breakup, and bifurcated streamers formation, as shown in Figure 15. Each of these is analyzed in detail.

### (a) Flow behind the liquid jet

As discussed above, the high and low-pressure region is created on the windward and leeward sides of the liquid jet. The pressure difference plays a prominent role in the deformation of the liquid jet cross-section, as shown in Figure 13. The pressure difference affects the flow passage around the liquid jet column. As shown on the mid-plane in Figure 16(a), the *high-pressure point* ($S_1$) is located at a particular height on the windward side of the jet where the static pressure reaches a maximum value. The streamlines using mean velocity with only x- and z-direction components are plotted in Figure 16(b) to estimate the flow features.

The low pressure in the liquid jet wake is dominant from the bottom wall at some height. It draws in the flow passing through the jet column's sides, creating a *counter-rotating vortex* (CRV), as shown by the yellow streamlines in Figure 16(b). Li et al. (2016) stated a saturation point on the leeward side of the liquid column. The recirculating flow may divide at the rear surface of the liquid column to create two circulation zones. In Figure 16(b), almost an entire leeward side of the jet column is exposed to only one (spanwise) vortical structure, i.e. CRV. The point $S_2$, where velocity would be zero on the leeward side of the liquid column, lies in the region of the complete breakup of the liquid column into ligaments.

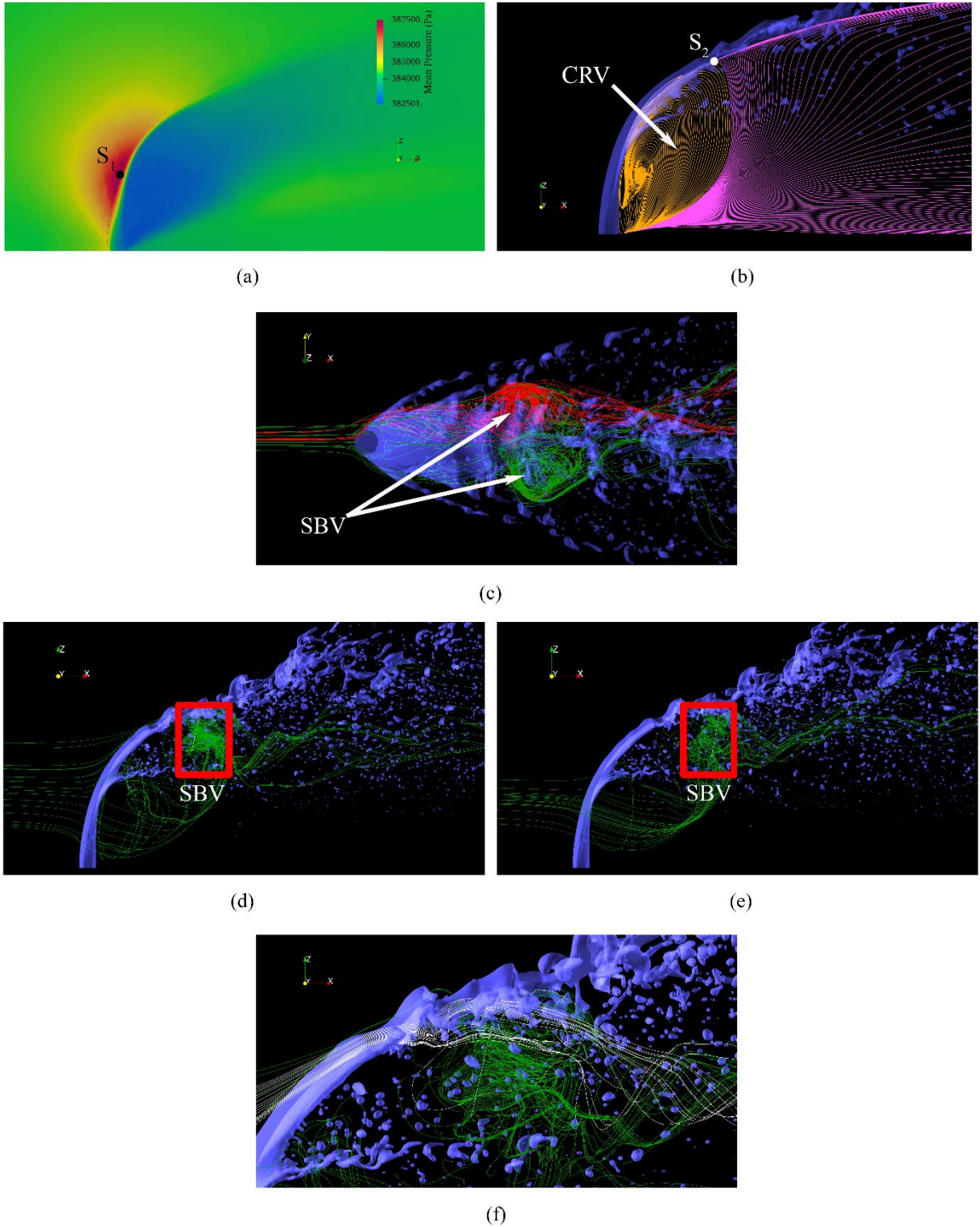

Fig. 16. (a) Pressure contour shows high-pressure point $S_1$ on the windward side of the liquid column, (c) Streamlines in x-z plane show counter-rotating vortex (CRV) in yellow color, (c) counter-rotating sheet breakup vortex (SBV) pairs (green-red color), (d) & (e) show movement of the SBV at consecutive time units following the bag of the wave. (f) The white-colored streamlines break through the liquid sheet bags, pulled by the SBVs for case 8 (q = 10 and We = 60).

Further, the two smaller counter-rotating vortex pairs act along the column breakup process downstream of CRV. An interesting phenomenon is noticed where these two vortices' centers move with the trough/bag of the liquid column instability. The movement of these vortices is shown at consecutive time units in Figures 16(d) and 16(e), where the vortex has shifted downstream following the trough of the wave. A closer look at this phenomenon in Figure 16(f) shows the 'white' colored streamlines passing through the ruptured liquid surface in the wave trough and mixing with outgoing 'green' streamlines of the vortex downstream.

Lower pressure at the vortex center pulls the flow-through ruptured liquid sheet. Wen et al. (2020) referred to these vortices as the *bag breakup vortex* (BBV). Here we refer to them as *sheet breakup vortex* (SBV) due to their role in the sheet-like liquid column breakup, which will be discussed in the subsequent section of the high shear case. The vortex strength of SBVs diminishes downstream as the liquid column disintegrates into ligaments and droplets. Though the streamlines corresponding to the mean velocity field do not show any vortex formation (picture not shown here), the movement of vortices downstream with diminishing strength indicates the vortex shedding phenomenon. Further, the vortex shedding phenomenon will be discussed later for the high momentum flux ratio case.

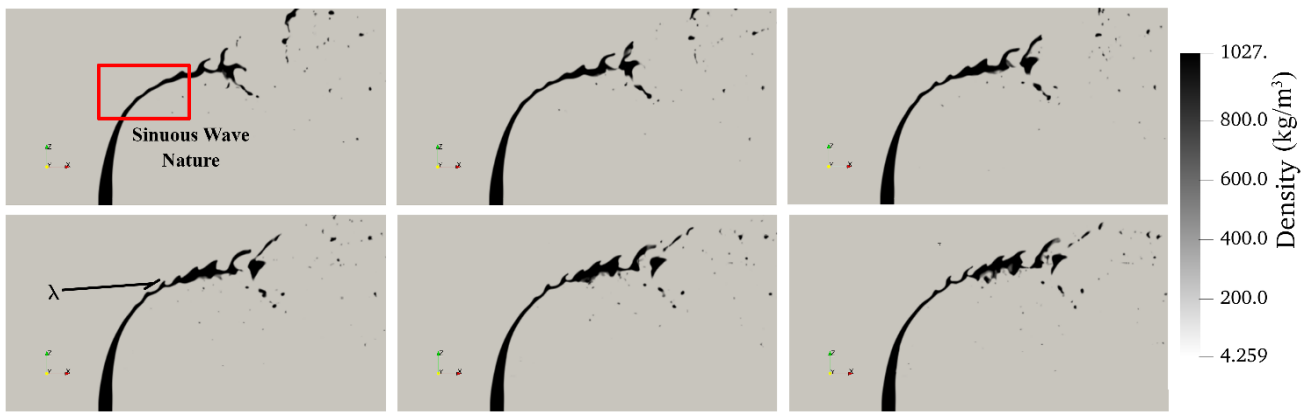

Fig. 17. Density contour at mid-plane of LJICF showing Kelvin-Helmholtz (KH) type instability. The typical KH type asymmetric waves are observed. The sinuous waves are most unstable here, considering the upper part of the liquid column behaves like a sheet (instead of varicose waves).

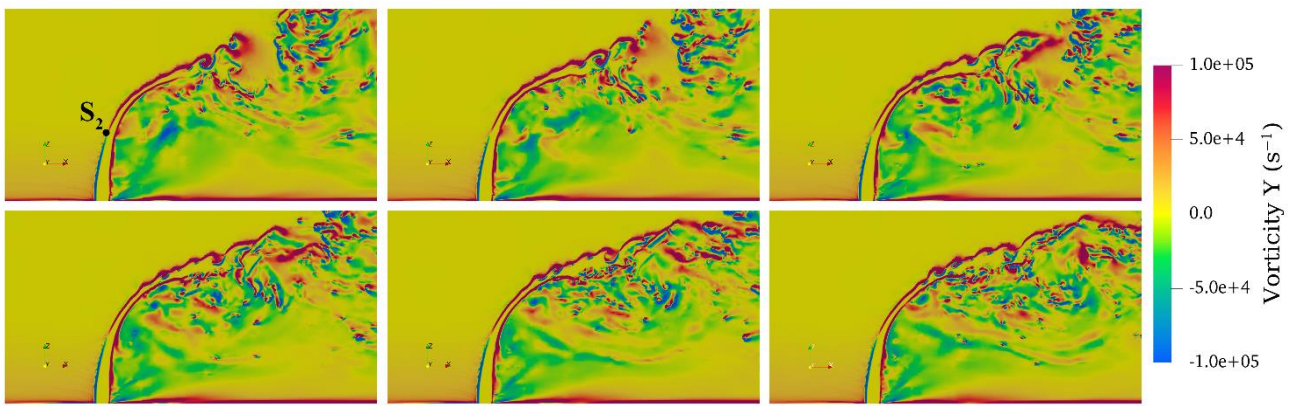

Fig. 18. The contour of Y-vorticity (normal to figure plain) at mid-plane of LJICF showing Kelvin-Helmholtz (KH) type instability. The near roll-ups are observed for this case.

**(b) Growth of instability on the windward side of the liquid in crossflow**

The formation of bags results from instability created upstream on the windward side of the liquid column. Earlier it was proposed that these are the Rayleigh-Taylor instability (Sallam, 2004; C.L.Ng et al., 2008)) is caused due to the higher/lower air pressure on the windward/leeward side of denser liquid which results in the bag formation similar to the finger/bag like structures obtained in the experiments of Lewis (1950), Taylor (1950). Rayleigh-Taylor instability is caused by denser fluid under acceleration towards lighter fluid.

In this case of lower $q$ and $We$, visual observation of the evolution of instability and breakup at consecutive time steps indicates that Kelvin-Helmholtz (KH) is the reason behind the instability growth (refer to Figure 17 showing density contour). Similarly, Y-vorticity (normal to the image plane) is also plotted to check the roll-ups of the two fluids (Figure 18). A small clip of the same has been provided as supplementary data, clearly showing the presence of KH-type instability.

As evident in Figure 19(a), the unstable waves are developed across the windward surface with parallel troughs/crests perpendicular to the stream direction. A cross-section of the liquid column is observed at a location just before the wave starts to develop, and it shows that the liquid column has transformed from a circular shape to that of a sheet (see Figure 19(a)). Hence, we treat this as a case of instability development on a liquid sheet. Also, the liquid sheet-like structure is located at a much downstream distance from the point $S_1$ along the liquid column such that high pressure has a lesser effect than the velocity shear between airflow and liquid column.

Considering inviscid, irrotational flow, the dispersion relation between wave number and wave growth rate is used for a liquid sheet as deduced by Squire (1953). If a coordinate system moves with the sheet interface at relative velocity $U$, an infinitesimal disturbance formed on it is described by:

$$\eta = \Re[\eta_0 \exp(ikx + \omega t)], \tag{36}$$

where $\eta_0$ is the initial wave amplitude, $k = 2\pi / \lambda$ is the wave number, and $\omega = \omega_r + i\omega_i$ is the complex growth rate. The dispersion relation for a moving liquid sheet under inviscid conditions can be derived as:

$$\omega^2 \left[\tanh(kh) + Q\right] + \omega 2iQkU - QU^2 k^2 + \frac{\sigma k^3}{\rho_1} = 0 \tag{37}$$

for the sinuous waves. Here, $Q$ is equal to $\rho_2/\rho_1$, $h$ is the sheet thickness. The solutions to the above Eq. 37 for the growth rate $\omega_r$ are:

$$\omega_r = \frac{\sqrt{\tanh(kh)QU^2k^2 - \sigma k^3/\rho_1\left[\tanh(kh) + Q\right]}}{\tanh(kh) + Q}. \tag{38}$$

Similar to above, Eqs. (37) and (38) corresponding to sinuous waves, the dispersion relation and the wave growth rate for the varicose mode of the waves are also obtained by replacing $\tanh(kh)$ by $\coth(kh)$ term in these equations.

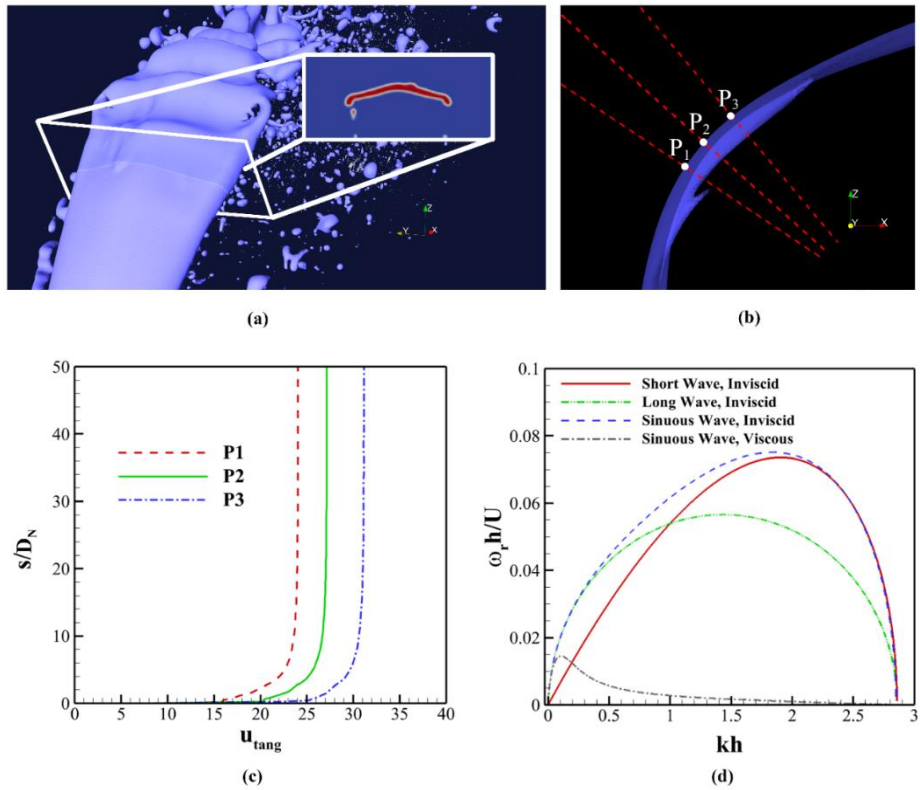

Fig. 19 (a) Location of a plane at which the cross-section of the liquid column is almost sheet-like, (b) three points showing the location curve for liquid profile calculation, (c) Velocity profile along the axis perpendicular to the interface, (d) Growth rate of different waves with wave numbers.

In the present case of LJICF, the velocities of the gas and liquid phase need to be parallel at the interface for the calculations. For this purpose, the velocity data is extracted along with the line segments at three points on the air-water interface, as depicted in Figure 19(b). The points are chosen at such a location where the instability waves start to appear. The line segments are perpendicular to the interface at their respective interface points. The magnitude of the velocity component is found along the line segment by $\left|\mathbf{u_{tang}}\right| = \left|\mathbf{u} \cdot \mathbf{n_t}\right|$, where $\mathbf{n_t}$ is the unit direction vector tangent to the surface. This provides

us with the velocity profile at desired points (shown in Figure 19(c)) and helps us calculate accurate velocity differences by avoiding the shear region. It is also visible that the maximum velocity increases from P1 to P3. Therefore, an average of the velocity difference $\Delta U_{avg} (\approx 17.56 \, m/s)$ is used in our calculation. Similarly, an average liquid core thickness $h_{avg} (\approx 1.5 \times 10^{-4})$ is also considered because of its variation along the liquid column, minimum at P3 and maximum at P1. The other details of thermophysical variables regarding the calculation are provided in Table. 1.

Senecal et al. (1999) discussed that the maximum growth rate of sinuous waves will always be greater or equal to the maximum growth rate of varicose waves in the moving liquid sheet. From the observation of our present case (refer to Figure 17, Figure 18), the sheet-like liquid column shows sinuous wave formation at its inception. Hence, we consider only the sinuous solution for the linear stability analysis (LSA) in this work. Figure 19(d) compares the growth rate for sinuous wave solution with short and long wave assumptions. Since the wavelength $\lambda$ of the disturbance in the present case is slightly smaller than $2\pi$ times the sheet thickness $h$, short waves start to dominate over long waves. This is evident in Figure 19(d) as well, where the short wave assumption (red curve) predicts near to the general inviscid sinuous mode growth rate (blue) as compared to the long-wave assumption (red curve). The long wave assumes $\tan(kh) \approx kh$ whereas the short wave assumes $\tan(kh) \approx 1$ in Eq. 38. Senecal et al. (1999) showed the dominance of long wave and short wave in low velocity/Weber number and high velocity/Weber number liquid sheets cases, respectively. The Weber number for the sheet-like liquid core is approximately $\approx 2.8$ with the present parameters, which makes it a marginal case for the distinction of long and short waves, albeit short wave is observed to dominate over a long wave in the present situation (as shown in Figure 19(d) and also, the assumption of $\tan(kh) \approx kh$ not holding true).

The $\lambda / D_N$ values from various correlations and linear stability analysis are compared with the computationally observed result in Table 8. The data corresponding to correlations and Rayleigh-Taylor breakup are calculated using the free stream air velocity of 41 m/s. The LSA error by Chandrasekhar and the general sinuous inviscid wave is minimum. The solution of a sinuous wave growth for a viscous sheet is very high, requiring further work to include viscosity effects. The Rayleigh Taylor or RT-based correlations (Chandrasekhar, 1961; C.L. Ng et al., 2008) predict almost 50% higher values confirming the absence of this type of breakup. This also explains the formation of cross-stream ligaments from the column breakup, as shown in Figure 19(a).

This case shows the development of KH instability and not the RT instability, which was considered the only dominant instability responsible for the column breakup in the past. Li et al. (2016) proposed that the liquid jet in very low crossflow velocity behaves similar to a jet in a quiescent air/gas medium and tends to develop KH instability. This is in contrast to the condition here. In the present case, air velocity is significant enough, and the momentum flux ratio is also low.

An essential factor, in this case, is the high shear experienced by the liquid column in the direction of liquid flow along the portion of its length where these KH waves start to develop. The high shear forces along the liquid column seem only possible when the momentum of crossflow air needs to be substantial compared to the momentum of liquid. The higher momentum of crossflow air bends the liquid jet acutely to high angles, which has a two-fold effect. First, the high-pressure region on the windward side of the liquid column is limited to the smaller area only, which faces directly into the incoming crossflow. Second, high velocity is achieved in the region beyond point $S_1$. Point $S_1$ lies at the approximately same area where the vorticity changes its sign as visible in Figure 18; the air velocity now exceeds the liquid flow velocity beyond point $S_1$. Thus, the high momentum of crossflow compared to the liquid momentum (that is, low momentum flux ratio) may be considered one of the governing reasons for the domination of KH instability. This is in contrast to Li et al. (2016). But we will see in the following sub-sections that it is not the crossflow velocity but the momentum flux ratio that decides the type of dominant instability.

| Correlations/LSA | $\lambda / D_N$ | Prediction error (%) |
|---|---|---|
| **KH – Short sinuous wave** | **0.86** | **25.75** |
| **KH – General sinuous wave** | **0.89** | **22.57** |
| KH – Sinuous wave, viscous - Senecal et al. (1999) | 15.05 | 1200.0 |
| **KH – Chandrasekhar (1961)** | **0.87** | **23.94** |
| Rayleigh-Taylor – Chandrasekhar (1961) | 1.73 | 50.00 |
| C.L. Ng et al. correlation (2008) | 1.83 | 59.09 |
| Sallam et al. correlation (2006) | 0.54 | 53.03 |
| **Present Simulation** | **1.15** | **-** |

Table 8. Comparison of wavelength observed in the present case and the predictions from LSA and correlations.

## (c) Bifurcation of liquid jet into streamers

In Figure 15 and also in Figure 20(a), we can observe two thread-like structures separating from the sides of the liquid jet column known as *streamers*. This distinct phenomenon is termed bifurcation and was observed in the experimental work of Sedarsky et al. (2010) and then in a computational study by Wen et al. (2020). Still, many researchers did not clearly explain the cause of this phenomenon. Sedarsky et al. (2010) suggested that it is due to the vortex formation behind the liquid jet column, and Wen et al. (2020) indicated that it could be due to the sheet break-up vortex (SBV). This streamer/bifurcation is formed in the near nozzle region close to the wall. As the jet penetrates more into crossflow (away from the wall), the liquid-gas circumference region is subjected to more shear forces. As discussed earlier, the boundary layer is formed on either side of the interface in both phases. An azimuthal instability develops across the liquid column, which may be responsible for the perturbations at the liquid column base. The cause of this has been reported to be Centrifugal Rayleigh-Taylor (CRT) instability (Behzad et al. (2015)). These instabilities grow along the liquid column axis along the internal boundary layer, resulting in the thickening of liquid column edges.

Data is extracted on two planes along the liquid column at locations just before and after the bifurcation – $C_1$ and $C_2$, respectively, as shown in Figure 20(a). Figure 20(b) shows the cross-stream (Z) component of vorticity and pressure on $C_1$ planes. Apart from the shear layer vorticity, the highlighted region indicates the presence of vorticity within the liquid, along the two peripheral edges. The vorticity is very weak compared to the boundary layer (for both air and liquid). Similarly, the pressure contour reaches a very low value at the center of these areas, as seen in Figure 20(b). It shows the continuous movement of liquid into its edges, which stretches the central region of the circular liquid core cross-section into a thin sheet-like structure while at the same time thickening these liquid column edges. Figure 20(c) presents the vorticity and pressure contours just after the detachment of the streamers on the $C_2$ plane, where these vortices on the liquid column periphery are absent. The separated boundary layer of the liquid column and bifurcations now passes through their gap (highlighted green). Also, the minimum pressure values are present in the gas phase only, outside the liquid.

As described in the previous section, the two factors play a significant role in the deformation of the cross-section of the liquid column – the internal boundary layer and the pressure difference driving the flow. The transfer of liquid to the periphery by the internal boundary layer in the absence of a return flow to the middle is majorly responsible for the thinning of the liquid column at the middle and the thickening of the edges.

Another important factor is the flow orientation of CRV, as shown by the velocity vectors and white-colored streamlines in Figure 20(d). A portion of the flow enters the wake of the liquid column from the bottom and gets drawn up by the CRV, and without reaching the bottom again, it exits the CRV from the sides. This sideward movement of air on either side of the liquid column pushes against the thick edges of the liquid column. This is also evident on the $C_1$ plane in Figure 20(b), labeled as CRV-exit, the thick opposite vorticity witnessed at the peripheries.

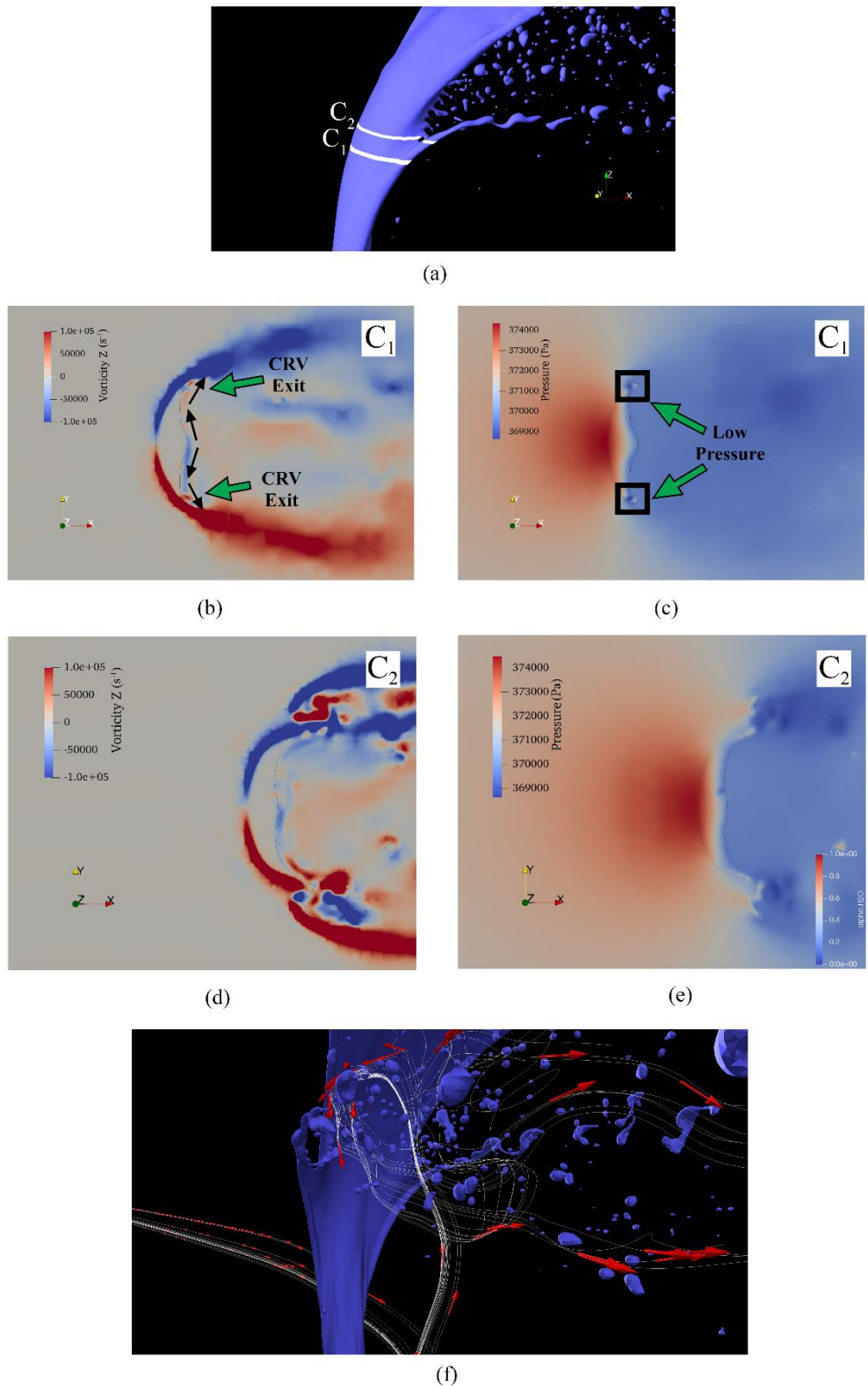

(a)

(b)

(c)

(d)

(e)

(f)

Fig. 20 (a) The location of planes – $C_1$ and $C_2$ on the liquid column just below and above the point of bifurcation, (b) Plane $C_1$ – Left column: Vorticity Z contours show weak vorticity within the liquid at edges (at the tip of thick green arrows and beside the internal boundary layer), whereas the black smaller arrows show the direction of flow along the edges; Right column: The pressure plot showing a drop at transverse liquid column edges, (c) Plane $C_2$ – Left column: Vorticity Z contours after bifurcation with negligible vorticity within it; Right column: Low-pressure area vanishes from the liquid edges just after the bifurcation; (d) The streamlines with stream vectors show the role of fluid exit from CRV in causing bifurcation/streamer formations.

Figure 20(c) shows changes in the vorticity just after separation (on the $C_2$ plane). The streamer is found to follow this fluid exiting the CRV, thus, following a different trajectory than a liquid column. Hence, it may be said that both the factors need to be present for a bifurcation to happen – moderate but sufficient level of shear flow/boundary layer to thicken the liquid column edges and the typical flow of CRV to separate and pull it along, away from the liquid column.

Also, Figure 15 shows two similar ligament-type structures alongside the bags of the liquid column at the top. This essentially re-thickens the edges because of the boundary flow within the liquid column. Hence, this phenomenon cannot occur because of shear instability but the internal boundary layer. The bifurcation formed is found to be thicker for low penetration cases. The bifurcation starts vanishing as the momentum flux ratio (q) increases (cases 1 to 4). This is because the primary break-up mode shifts towards pure shear mode as the momentum flux ratio or crossflow Weber number increases, enhancing the surface stripping process with crossflow. In cases 4, 9, and 10, where the momentum flux ratio is high, we have observed multiple streamer formations in which the liquid jet column is split into multiple(four) membranes. The bifurcation formed is very thin and can be considered almost nil in these cases.

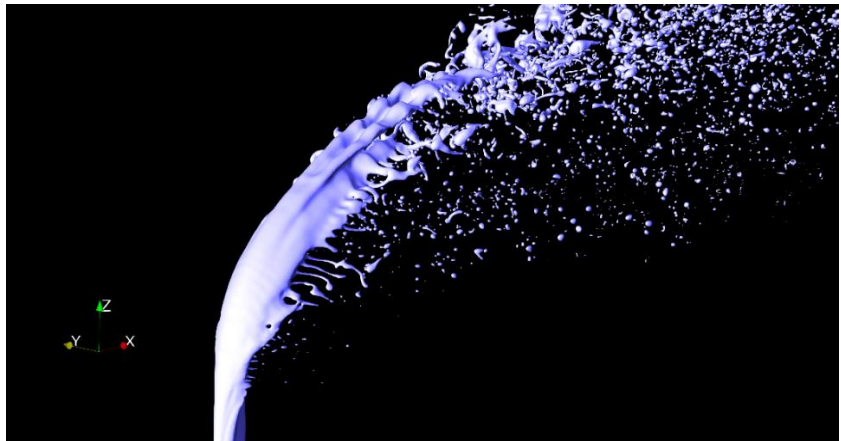
Fig. 21. Basic figure of case 5 with We=150 and q=10.

### 3.3.2 Shear Regime

The LJICF shows a shear-dominated breakup at a high Weber number but a lower momentum flux ratio in Figure 21. Case 5 out of the two high We cases is chosen for the study. The q value remains the same as the previously analyzed case 8. Here, the flow feature behind the LJICF and the instability at the edges are discussed.

**(a)    Flow behind the liquid column**

The flow feature behind the liquid jet column for We=150 (Figure 21) is similar to the previous case with We=60. The pressure contour (Figure 22(a)) is similar to the low We case. In this case, only a single recirculation zone is visible, as shown by 'yellow' colored streamlines in Figure 22(b). The saturation point of the leeward side of liquid column $S_2$ lies in the region where the complete breakup of the liquid column takes place. Similarly, the liquid shear at the edges vanishes downstream of point $S_2$. The shear breakup will be covered below in Section 3.2.2(b). The two opposite rotating vortex pairs (referred to as SBV in the previous section) are also present, as shown in Figure 22(c), albeit their center positions shift one behind the other, and one of them is weaker than the other. In Figure 22(c), the red streamlines pass through the thin liquid sheets as it tears apart. The liquid sheet breakup is delayed on the side of the weaker 'green' colored vortex. Hence, these vortices play an essential role in the breakup of thin liquid sheets present on both sides of the thicker liquid column. These liquid sheets break up without forming a bag-type hollow structure; hence these vortices' name '*sheet breakup vortex*' (SBV).

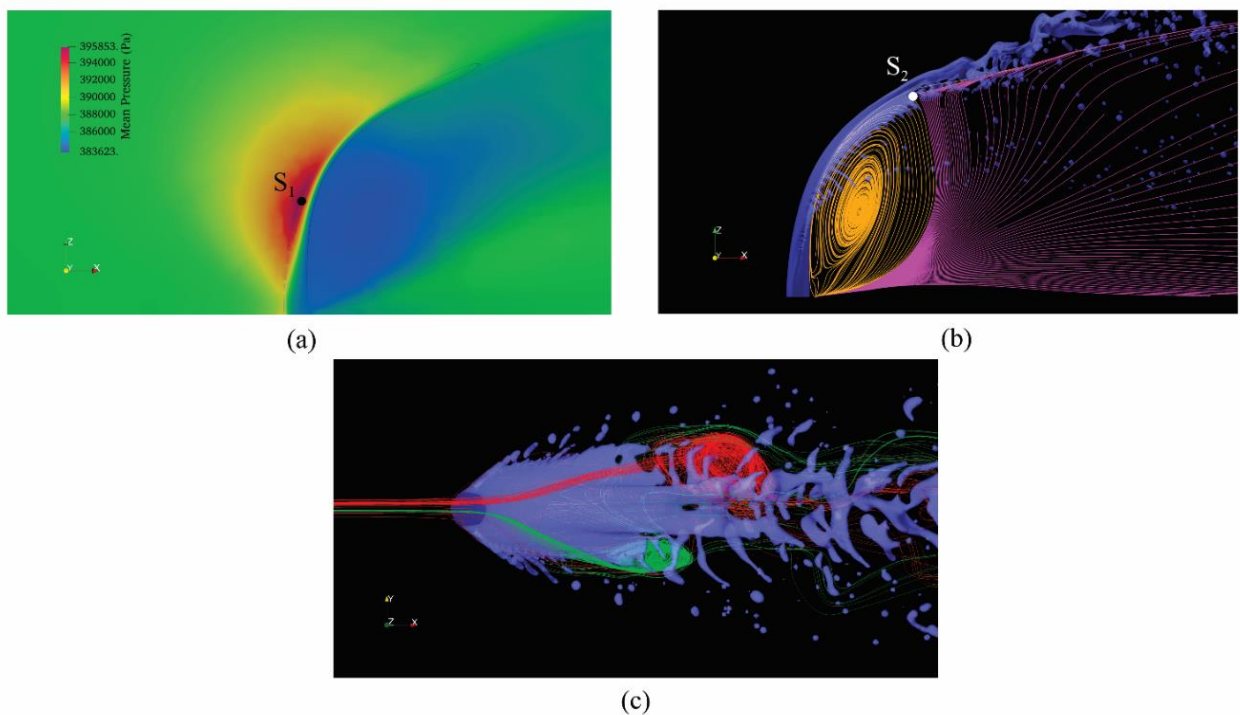

Fig. 22. (a) Pressure plot with stagnation point $S_1$, (b) The mean flow on mid-plane showing same flow as the first case with leeward stagnation point $S_2$, (c) The two counter-rotating SBVs for case 5 (q = 10, We = 150).

### (b) Growth of instability and Streamers

The two different disturbances are apparent here – one at the end liquid column and the other along the edge of the liquid column. The density contours at the middle plane (XY), bisecting through the liquid column, were analyzed for the first type of disturbance. It is not further analyzed here since the breakup happens just after the inception of the not-so-prominent first or second wave or even before that. Instead, a more dominant shear breakup is focused on.

As shown in Figure 23, the two different wavelength waves appear on the edges of the liquid jet. The waves on the upper side are almost double the wavelength at the bottom of the liquid column, increasing along the liquid column. Another aspect is the formation of ligaments and subsequent thick droplets in the place of bifurcation/streamer. The region of a larger wavelength lies just above this bifurcation point 'B'.

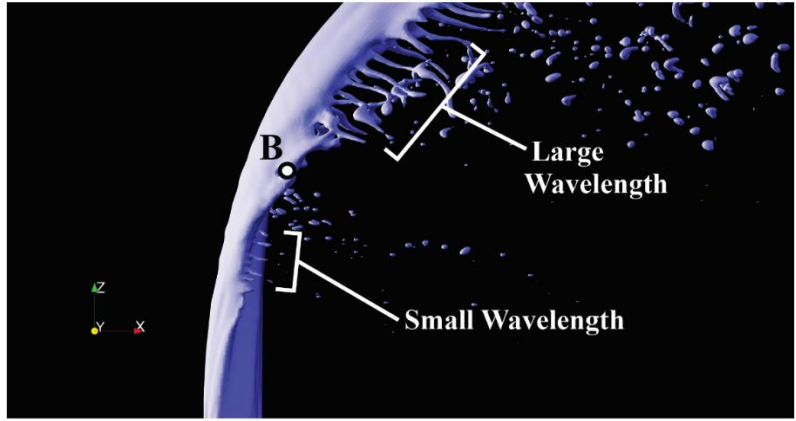

Fig. 23. Shear instability waves at liquid-core edges with different wavelengths at the top and bottom of point B.

The C.L. Ng et al. (2008) correlation from their experiment shows the dependence of $\lambda/D_N = 4.3We_G^{-0.33}$ which predicts almost four times the wavelength observed near the bottom and almost double the wavelength above point B. The different behavior witnessed in the two regions may be due to the change in the size of the liquid column faced by the incoming air; that is, the bottom part of the liquid column is analogous to the circular cylinder of size same as the jet diameter whereas, the upper part (above point B) is similar to a shell of size larger than jet diameter. Considering the disturbances in the lower portion, there is a similarity with the instabilities in separating shear flow over the cylinder with $Re > 1000$. At a high Reynolds number case like this, two kinds of 3D instabilities exist – a) streamwise vortices of separating shear layer and b) streamwise vortices of the wake. In the present study, the instabilities of separating the shear layer are more critical in assessing its effect on the liquid edges. Based on the results of Bernal and Roshko (1987), Williamson (1996) proposed the following relation for the wavelength of streamwise vortices of separating shear layer:

$$\lambda_{SL}/D_N \approx 25\,Re^{-0.5}.$$

(39)

The Re corresponds to the Reynolds number of airflow around the circular liquid column, which behaves like a cylinder of diameter $(D_N) \approx 8910$. The prediction is near the wavelength obtained from simulation for the lower part of the liquid jet column, as shown in Table 9 for both case 5 and case 6 with the same *We*, though there is some deviation for case 9. It also shows that the prediction is independent of the momentum flux ratio; it does not depend on the liquid jet velocity. Similarly, the study of instability in the upper part of the jet column can be part of future work concerning the flow structures over hollow shells.

| Correlations | $\lambda / D_N$ | | |
| --- | --- | --- | --- |
| | **Case 5**<br>(q=10, We=150) | **Case 6**<br>(q=19, We=150) | **Case 9**<br>(q=77, We=60) |
| **Williamson correlation (1996)\*\*** | **0.26** | **0.26** | **0.33** |
| C.L. Ng et al. correlation(2008) | 0.82 | 0.82 | 1.11 |
| **Present Simulation** | **0.21** | **0.26** | **0.21** |

Table 9. Comparison of wavelength observed in this study with the correlations. \*\*Williamson correlation about shear layer streamwise instability is valid, assuming that the same instability triggers the shear instability in liquid edges.

### 3.3.3 Surface Breakup Regime

Case 9 with the same Weber number 60 as case 8 but higher momentum flux ratio of 77 is chosen in this regime.

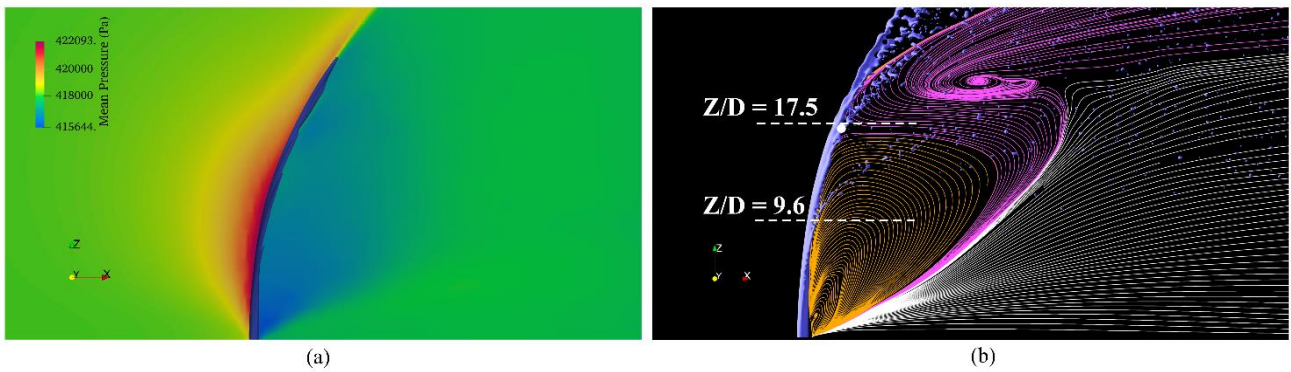

Fig. 24. (a) Pressure plot with longer and extended high-pressure zone on the windward side of LJICF, (b) Streamlines showing two recirculation regions separated with a stagnation point S$_2$. The two horizontal planes are the location for the streamlined contours shown in Figure 25.

### (a)     Flow field behind the liquid jet column

In this case, the flow is quite distinct from the low momentum flux ratio cases, as shown in Figure 24. In Figure 24(a), there is an extended high-pressure region on the windward side of the liquid column with a higher value of the maximum pressure. The existence of high pressure for most of the windward side of the liquid owes to deeper penetration by this liquid jet. The streamlines on a plane shows two recirculation zones ('yellow' colored vortex as CRV and 'pink' colored vortex lies above CRV) of opposite rotations in Figure 24(b), different from previous cases. From this figure, the saturation point can easily be found on the leeward side of the liquid jet. It is the same as the observation made by Li et al. (2016) in their case with the momentum flux ratio of 88.2, which was kept constant for all their three simulations. The variation of the Weber number was only employed. Hence the condition of the present case nearly matches the conditions used by Li et al. (2016).

Figure 25 represents the instantaneous streamlines on horizontal planes at two planes – $Z/D_N$= 9.6 (Figures 25(a) – 25(f)) and 17.5 (Figures 25(g) – 25(l)) against the density contour, which helps to locate flow features concerning the cross-section of the liquid column (green colored). The location of planes is chosen appropriately, considering that it lies near the point of bifurcations (or streamer formations), as shown in Figure 24(b). This allows us to observe the effect of bifurcated streamers on the flowfield, if any. First, the streamlines at plane $Z/D_N$ =9.6 depict periodic vortex shedding. This resembles the flow over a cylinder or, more appropriately, over the semi-spherical shell (a shape similar to the liquid jet-core cross-section in Figures 25(a)-25(f)). One of the vortices in Figure 25(c) (shown by red pointer) undergoes vortex tearing in Figure 25(d), during which one of them remains near the jet column while the other convects downstream with the flow. Second, Figures 25(g)-25(l), on the right side column, captures the bifurcation of the second streamer as shown by the green-colored stretched liquid column at $Z/D_N$ =17.5 and the vortex shedding near its location on the leeward side. The vortices at $Z/D_N$ =17.5 (right) are stronger and bigger than that observed at $Z/D_N$ =9.6 (left), and thus, they are present for a longer downstream distance. Another feature is the vortex pairing, occurring downstream of the liquid column in Figures 25(j) – VP-1 and VP-2. The two vortices prior to a pairing process are shown in Figure 25(h)-(i). After completion of the pairing process in Figure 25(j), they give rise to a stronger vortex, as observed in Figure 25(k).

The vortex pairing on the leeward side seems related to the liquid core and its bifurcated streamer formations. The streamers' formation results in the gap between the central liquid core and streamers, affecting the vortex shedding downstream. It may be said that the different vortices shed in the bifurcation process interacts with each other leading to the vortex pairing (as noticed in Figure 25(h)-(j)). It further complicates the flow concerning the interaction between liquid core breakup into streamers and the airflow development behind it, both dependent on each other.

Since the crossflow air velocity remains the same as for the low momentum flux ratio case in the bag breakup regime (Section 3.2.1), the vortices witnessed (SRVs) resulting from vortex shedding are similar to vortices on plane $Z/D_N$ = 9.6. However, the strength of these vortices will vary for both these cases because of the difference in the trajectory of the liquid jet, size, and strength of CRV. Since the SRVs are found to move along the KH waves in case 8, it may be an interesting future study to confirm if it is always true.

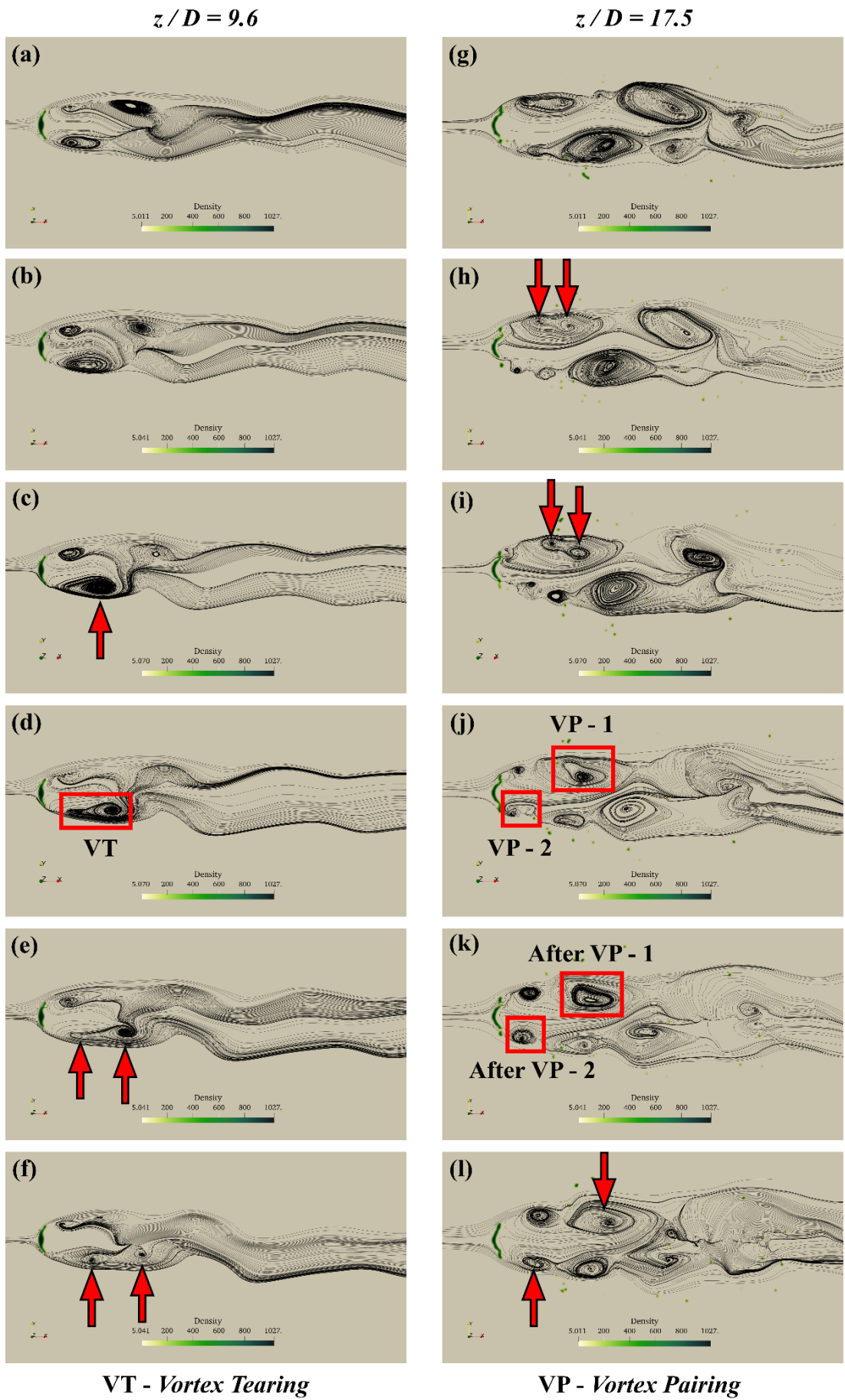

Fig. 25. Left panel: Streamlines show vortex shedding along with a vortex tearing phenomena on plane Z = 9.6; Right panel: Streamlines show vortex shedding and (two) pairing processes on plane Z = 17.5.

**(b)     Instability on the liquid column and surface edges**

This case shows both the instabilities of – the liquid column and its edges. The Rayleigh-Taylor instability is found to be dominant in the liquid column. The waves are symmetric until the breakup, and the typical Kelvin-Helmholtz roll-ups are absent, as shown in Figure 26. The wavelength can be predicted by using C.L.Ng et al.'s (2008) assumption of cylinder drag as acceleration for Rayleigh Taylor instability in this case:

$$C_{D,cylinder} \approx 1 + 10 \text{Re}_G^{-2/3} \qquad \text{when} \quad 1 < \text{Re}_G < 2 \times 10^5. \qquad (40)$$

and,

$$\lambda / D_N = 5.3 \text{We}_G^{-0.26}. \qquad (41)$$

Here $C_D$ is the coefficient of drag for the cylinder and $\text{Re}_G$ is the air Reynolds number. The wavelength observed in the present case is compared with the LSA result by Chandrasekhar (1961) and correlation by C.L.Ng et al. (2008) in Table 10.

This case with high momentum flux ratio further confirms the claim that as the momentum flux ratio increases, the chances of Rayleigh-Taylor dominated instability is higher. In the case of high momentum flux ratio and moderate (or low) Weber number, the liquid momentum is higher than gas/air momentum. In other words, the injected liquid/water mass flow rate is also high, unlike case 5, where the liquid jet bends acutely due to early flattening. It leads to deeper penetration of liquid jet with almost vertical liquid jet through the crossflow. The portion of the liquid column that faces directly into the incoming crossflow air has an extended high-pressure region. At the same time, the air/gas cannot generate enough shear along the liquid flow direction. Hence, the effect of high pressure on the windward side starts to have its outcome in Rayleigh-Taylor instability.

| Correlations/LSA | $\lambda / D_N$ |
|---|---|
| Rayleigh-Taylor - Chandrasekhar | 1.73 |
| C.L. Ng et al. correlation (2008) | 1.83 |
| **Present Simulation** | **1.70** |

Table 10. Comparison of wavelength observed in the present case and the predictions from LSA and correlations.

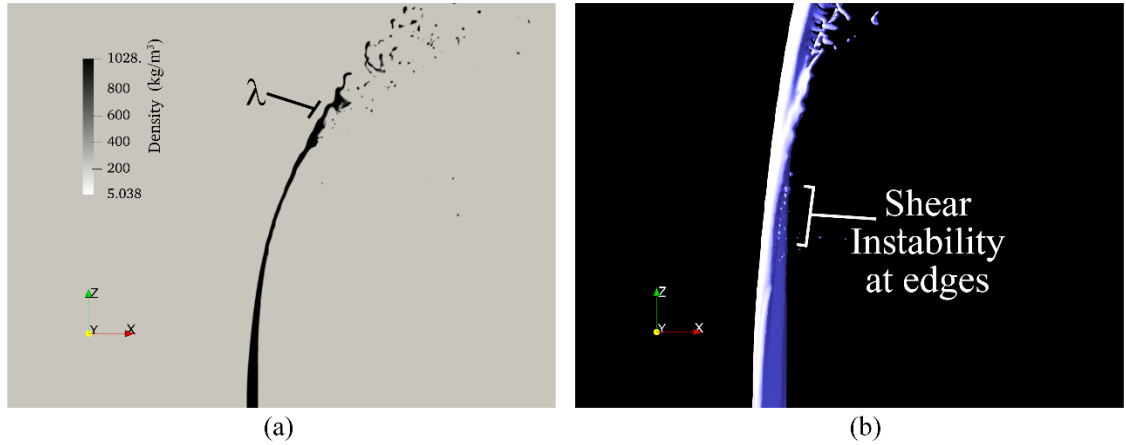

Fig. 26. (a) Density contour at mid-section of LJICF show Rayleigh-Taylor instability, (b) Shear instability for We = 10 and high momentum flux ratio (q) = 77.

It is, thus, proposed here that the momentum flux ratio plays a vital role in deciding the type of instability growth on the liquid column. The momentum flux ratio is lower, higher the probability of developing KH type of instability. Since the present study primarily focuses on the rigorous validation of compressible solver against varying parameters, further detailed analysis at varying q and We is beyond the scope of this work.

The wavelength of surface waves along the transverse edges of the liquid column has been calculated using the assumption discussed in the previous section. It is proposed that the same disturbance causes this instability of a liquid column as that of the air shear layer around a cylinder. Eq. 39 is used for the wavelength calculations in Table 9. Using this assumption, the Williamson shear layer correlation predicts $\lambda / D_N$ equals 0.33, which is closer to the observed (non-dimensionalized) wavelength of 0.26 from the simulations compared to 1.18 by C.L.Ng et al. (2008)'s correlation. C.L.Ng et al. (2008) correlation may be based on the wavelength measurements downstream of the liquid column for cases with much higher q. However, the empirical correlation by Williamson is only valid for the region near the injection nozzle where the cross-section of the liquid column is nearly circular.

**(c)    Bifurcations**

The bifurcation increased to four but thinner and a little vaguely visible (compared to case 8) at a higher momentum flux ratio, as shown in Figure 24(b). The parameters We (equal to 60) and q (equal to 77) of this case are similar to case 3 (We = 68, q = 64) of Sedarsky et al. (2010), where they also observed multiple streamers. This may be owed to the two circulation zones witnessed in the mean flow behind the liquid core (Figure 27). Following the same explanation of Section 3.3.1(c), the CRV exit carries along and diverts the first bifurcation/streamer liquid. Similar to the previous case, the fluid exit also happens from the sides of these recirculation zones formed on the leeward side. As discussed earlier, the flow is highly complex, with a mix of vortex shedding and recirculation. The vortex shedding induces the sideward flow of air near the leeward side of the liquid column, which in turn pushes against and separates the thick transverse edge from the liquid column. Hence, in this case, the two crucial factors are the thickening of the edge by the internal boundary (shear) layer and the separation of streamers from the central liquid column by the CRV/recirculation flow.

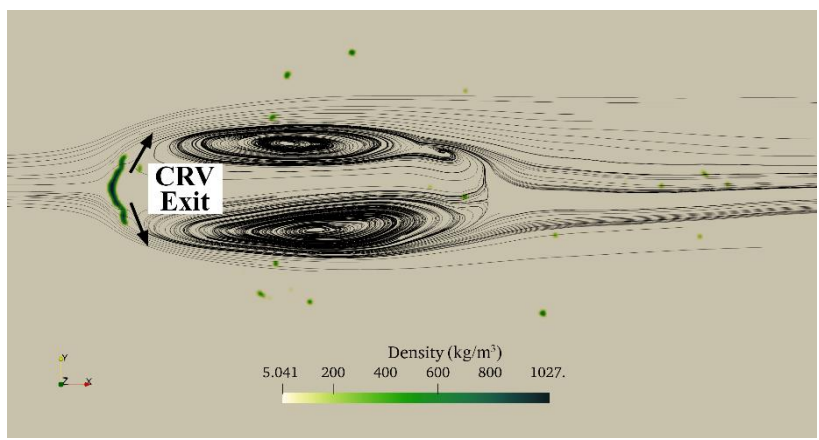

Fig. 27. Streamlines show mean flow on plane Z/D=17.5. The clockwise rotation of the bottom vortex and anti-clockwise of the upper vortex show the outward movement of gas/air near the leeward side of the liquid column. This sideward movement is responsible for streamer formation.

## 4. Conclusion

This work numerically simulates a liquid jet's primary and secondary atomization in crossflow using a compressible Volume of Fluid (VOF) - Lagrangian Particle Tracking (LPT) coupled solver implemented in OpenFOAM. The iso-Advector scheme by Roenby et al. (2016) is used to capture the droplets and sub-grid fluid distribution. In contrast, the coupling algorithm by Heinrich and Schwarze (2020) provides flexibility in demarcating the droplets and converting them into Lagrangian particles under satisfying conditions. The complete framework is validated against the comprehensive experimental data of Amighi (2015). The numerically predicted data for liquid jet penetration, droplet size characteristics $D_{32}$, and STD agrees with experimental data. The effect of various parameters, namely liquid jet velocity, crossflow velocity, and pressure, on the droplet size characteristics also predicts similar trends as in the experiments. The comparison of liquid jet penetration with empirical correlations shows that the predicted trajectory is closest to Elshamy's correlation, and correlations based on momentum flux ratio alone are found to overpredict by a large margin. In line with the previous literature, the comparison of the stream-wise location of the breakup for each case is constant at $9.2D_N \pm 1.2D_N$, independent of the momentum flux ratio.

Few cases are analyzed in different breakup regimes. The momentum flux ratio is a governing factor for the instability that dominates the liquid column flow. The Kelvin-Helmholtz type instability causes the liquid column to break up at a lower momentum flux ratio. The crossflow air momentum forces the liquid column to bend to sharper angles, producing a higher shear force along the liquid flow direction. Based on the observations of a sheet-like cross-section in the latter portion of the liquid column, the inviscid and viscous linear stability analysis results are computed considering the thickness of the sheet-like liquid column. It is found that the sheet Weber number is high enough so that only the short wavelength instability dominates the breakup. The results are compared to the wavelength detected from the simulations.

On the other hand, the Rayleigh-Taylor instability is dominant for the high momentum flux ratio case. It is inferred that high momentum flux ratio causes high penetration, which is a reason for the extended high-pressure zone on the windward side of the liquid column. The increased air pressure and lesser shear results in the growth of Rayleigh-Taylor-type instability. The wavelength from the simulations closely matches with linear stability analysis and correlations.

The shear instability along the transverse edges of the liquid column is well captured in the simulations. It is hypothesized that the instability of the liquid column at the edges is caused due to the instability of the air shear layer passing by around the liquid column. The wavelength measured at the bottom part of the liquid column from simulations was compared to Williamson's empirical correlation of shear layer instability for a flow over the cylinder (1996). It is closer to the simulated values than the correlation results from the past literature.

The bifurcation or streamer generation, evident at lower to moderate values of momentum flux ratio and Weber number, is another crucial aspect that has been observed. These bifurcations are caused by the counter-rotating vortex's three-dimensional structure and internal boundary layer shear at the liquid jet's windward side. The shear breakup at the edges causes the streamers to be mostly non-existent or thin at a higher momentum flux ratio or Weber number.

To summarize, LJICF is a classical problem that involves complex flow physics. The present work attempts to develop and access an accurate, robust platform (hybrid compressible VOF-LPT framework) while investigating the break-up phenomenon in LJICF. The other aspects, like break-up regimes and details of instability behavior, are ongoing work in the same group and beyond the scope of the present study.

## Acknowledgments

Financial support for this research is provided through the Department of Science and Technology (DST) under National Supercomputing Mission (NSM), India. We acknowledge the National Supercomputing Mission (NSM) for providing computing resources of 'PARAM Sanganak' at IIT Kanpur, which is implemented by C-DAC and supported by the Ministry of Electronics and Information Technology (MeitY) and Department of Science and Technology (DST), Government of India. Also, we would like to thank the computer center (www.iitk.ac.in/cc) at IIT Kanpur for providing the resources to carry out this work.